\documentclass[%
 reprint,
 amsmath,amssymb,
 aps,
floatfix,
]{revtex4-2}

\usepackage{tikz}

\usepackage{graphicx}% Include figure files
\usepackage{dcolumn}% Align table columns on decimal point
\usepackage{bm}% bold math
\usepackage{appendix}
\usepackage{subfigure}
\usepackage{xcolor}
\begin{document}

\preprint{APS/123-QED}

\title{The subgrid scale pressure field of scale-enriched Large Eddy Simulations using Gabor modes}% Force line breaks with \\

\author{Ryan D. Hass}
 \affiliation{Department of Mechanical Engineering, Stanford University, Stanford, CA 94305}

\author{Aditya S. Ghate}
\affiliation{
 Department of Aeronautics and Astronautics, Stanford University, Stanford, CA 94305
}%

\author{Sanjiva K. Lele}
\affiliation{%
 Department of Mechanical Engineering
}%
\affiliation{Department of Aeronautics and Astronautics, Stanford University, Stanford, CA 94305}

\date{\today}% It is always \today, today,

\begin{abstract}
With the continuing progress in large eddy simulations (LES), and ever increasing computational resources, it is currently possible to numerically solve the time-dependent and anisotropic large scales of turbulence in a wide variety of flows. For some applications this large-scale resolution is satisfactory. However, a wide range of engineering problems involve flows at very large Reynolds numbers where the subgrid scale dynamics of a practical LES are critically important to design and yet are out of reach given the computational demands of solving the Navier Stokes equations; this difficulty is particularly relevant in wall-bounded turbulence where even the large-scales are often below the implied filter width of modest cost wall modeled LES (WMLES). In this paper we briefly introduce a scale enrichment procedure which leverages spatially and spectrally-localized Gabor modes. The method provides a physically consistent description of the small scale velocity field without solving the full non-linear equations. The enrichment procedure is appraised against its ability to predict small scale contributions to the pressure field. We find that the method accurately extrapolates the pressure spectrum and recovers pressure variance of the full field remarkably well when compared to a computationally expensive, highly resolved LES. The analysis is conducted both in \textit{a priori} and \textit{a posteriori} settings for the case of homogeneous isotropic turbulence (HIT).
\end{abstract}

\maketitle

%\tableofcontents

\section{Introduction}
There are many examples of fluid flows in engineering applications where the Reynolds number far exceeds the practical limit for numerically resolving all turbulence scales. One such example is the atmospheric boundary layer through a wind farm. Such simulations are of obvious interest given the global climate-energy problem and yet current understanding of atmospheric turbulence and its effect on wind farm performance and wind turbine life-expectancy is not well understood. In high Reynolds number cases such as this it is only practical to resolve scales much larger than the turbine blade chord length \cite{GhateThesis} thus limiting our understanding of fluctuating loads on critical components and preventing the ability to design accordingly. Similar constraints are faced in design analysis and optimization of aerospace systems, which typically use RANS-based CFD tools. As outlined in the NASA CFD Vision 2030 Study \cite{slotnick2014cfd}, RANS-based turbulence models have limited predictive accuracy when dealing with separated flows, complex flow interactions, etc. and there has been growing interest in methods which resolve some range of turbulence scales such as LES and hybrid RANS-LES \cite{ghate2020scale,spalart2015philosophies}.

Ghate \& Lele \cite{GhateJFM2017} developed a method to enrich turbulence scales below the implied filter width of LES in such a way that local information of subgrid-scale dynamics are accurately represented through the use of spatially- and spectrally-localized Gabor modes. The method relies on a \textit{quasi-homogeneous assumption} and represents the subgrid velocity field as a stochastic Fourier-Stieltjes integral according to Batchelor (1953) \cite{batchelor1953theory} in Eq. (\ref{Synthesized field}), where the stochastic $d\bm{Z}$ modes must attenuate beyond some pre-determined length scale proportional to the size of the quasi-homogeneous region. These spatially-localized stochastic modes are referred to as \textit{Gabor modes}. 

The spatial localization is provided by appeal to the Gabor transform (Eq. (\ref{Gabor Transform 2})) which amounts to a windowed Fourier transform where $f_{\epsilon}$ is a smooth, compact support window function whose width is controlled by the parameter $\epsilon$.

\begin{equation}\label{Synthesized field}
    \begin{aligned}
    \bm{u}_s(\bm{x},\bm{x}_0,\Delta_w) &= \int_{\bm{k} \in \mathbb{R}^3} e^{i\bm{k}\cdot \bm{x}} d\bm{Z}(\bm{k},\bm{x}_0,\Delta_w) \\
    \big\langle dZ_i(\bm{k}) dZ_j (\bm{k}') \big\rangle &= \delta(\bm{k} - \bm{k}') \Phi_{ij}(\bm{k})d\bm{k}
    \end{aligned}
\end{equation}

\begin{equation}\label{Gabor Transform 2}
    \widetilde{\bm{u}}(\bm{x}_0,\bm{k}_m,t) = \int_{\bm{x} \in \mathbb{R}^3}\bm{u}(\bm{x},t) f_{\epsilon}(\bm{x}-\bm{x}_0) e^{-i \bm{k}_m \cdot (\bm{x} - \bm{x}_0)}d\bm{x}
\end{equation}

\noindent Here, and in all subsequent equations, $\bm{x}$ is the spatial coordinate in a three dimensional Cartesian grid who's components are $x$, $y$, and $z$, or equivalently $x_i$ for $i\in\{1,2,3\}$; $\bm{k}$ is the corresponding wave-vector with components $k_i$; $\Phi_{ij}$ is the velocity spectrum tensor and is the Fourier transform of the two-point correlation $R_{ij}(\bm{r}) = \big\langle u_i(\bm{x}) u_j(\bm{x}+\bm{r}) \big\rangle$; $\bm{x}_0$ is the physical location of the Gabor mode; and $\Delta_w$ is the spatial extent, or region of influence, of the Gabor mode determined by its window function.

The physical restrictions on the size of the enriched sub-domains (formally referred to as \textit{quasi-homogeneous regions} \cite{GhateJFM2017}), and thereby the region of the domain influenced by Eq. (\ref{Synthesized field}), is bounded by two requirements: 1) it must be significantly larger than the largest scale to be enriched and 2) it must not be so large that the quasi-homogeneity assumption breaks down. Further details regarding the reconstruction of the small scale velocity field can be found in Ghate (2018) \cite{GhateThesis}.

Under this construct Ghate \& Lele (2017, 2020) \cite{GhateJFM2017,ghate2020broadband} have shown that scale enrichment is capable of predicting a physically accurate small-scale velocity field when compared to higher fidelity simulations. A summary of the method and key results are included in section \ref{Energetics} so that the background required for the present development is concisely available. Section \ref{Pressure} demonstrates the enrichment method's ability to represent the subgrid scale contribution to the pressure field in the context of incompressible homogeneous isotropic turbulence. Two test cases are considered in this section: statistically-steady (i.e. forced) HIT and decaying HIT. Both simulations mimic the $\text{Re} \to \infty$ limit by setting the molecular viscosity to zero. For comparison, results from a finite Re DNS are included in appendix \ref{Appendix - Finite Re DNS}. We end with some broader comments on the results and its extension.

\section{Summary of previous work}\label{Energetics}

Ghate \& Lele showed that a discrete approximation of Eq. ($\ref{Synthesized field}$) can accurately represent the subgrid velocity field given that the size of the quasi-homogeneous region is chosen to satisfy the conditions specified in the previous section. Thus the synthesized small-scale velocity field is given by

\begin{equation}\label{Synthesize field (discrete)}
    \bm{u}_s = \sum_j f_{\epsilon}(\bm{x} - \bm{x}_j) \bm{a}_j e^{i \bm{k}_j \cdot (\bm{x} - \bm{x}_j)}
\end{equation}

\noindent where $\bm{a}_j$ is a complex valued amplitude such that the synthesized velocity field is solenoidal and yields the correct energy spectrum. This representation is reminiscent of the kinematic simulations of Fung et. al \cite{fung1992kinematic} and is sufficient for predicting second order statistics of isotropic turbulence. To capture the effect of large-scale inhomogeneity prevalent in high Reynolds number wall-bounded shear flows we appeal to rapid distortion theory (RDT) where the local large scale velocity gradient is used as the mean shear for the Gabor modes. An RDT linearization is not valid in general however, due to the fact that turbulence time scales can be much smaller than the mean shear time scale in the log-law region of wall-bounded flows. As a result we appeal to J. Mann's eddy life time hypothesis \cite{mann1994spatial} where each mode is evolved according to the RDT equations for a $k$-dependent time horizon. Further details of the straining procedure and enrichment of wall-bounded flows are provided in \cite{GhateJFM2017} and \cite{GhateThesis}.

\begin{figure*}[htp]%[hbt!]
    \centering
	\subfigure[$32^3$ LES (upsampled)]{\includegraphics[width=.32\textwidth]{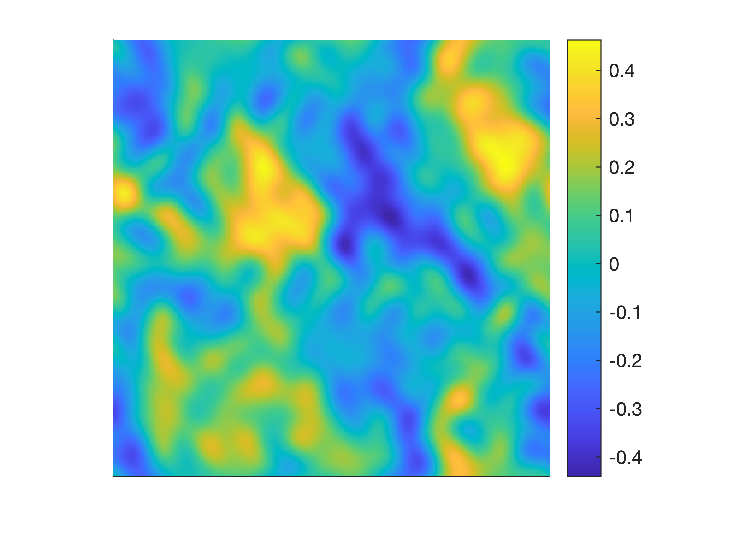}}
	\subfigure[Small scales (Gabor modes)]{\includegraphics[width=.32\textwidth]{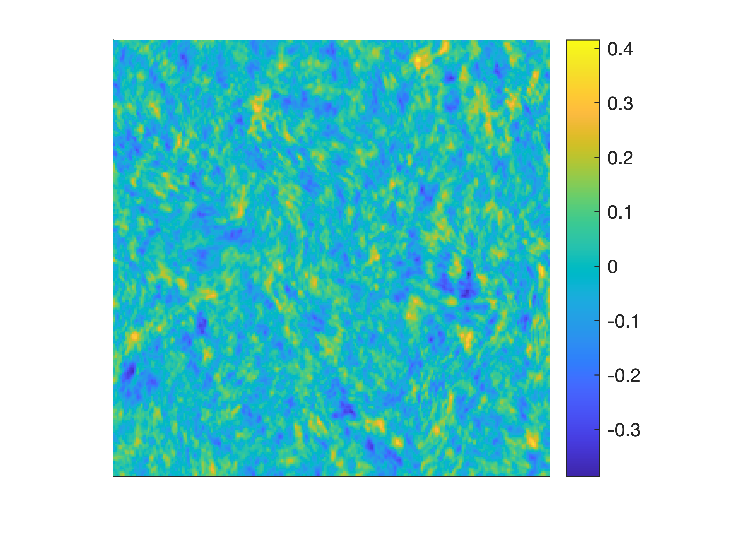}}
	\subfigure[Full/Enriched field]{\includegraphics[width=.32\textwidth]{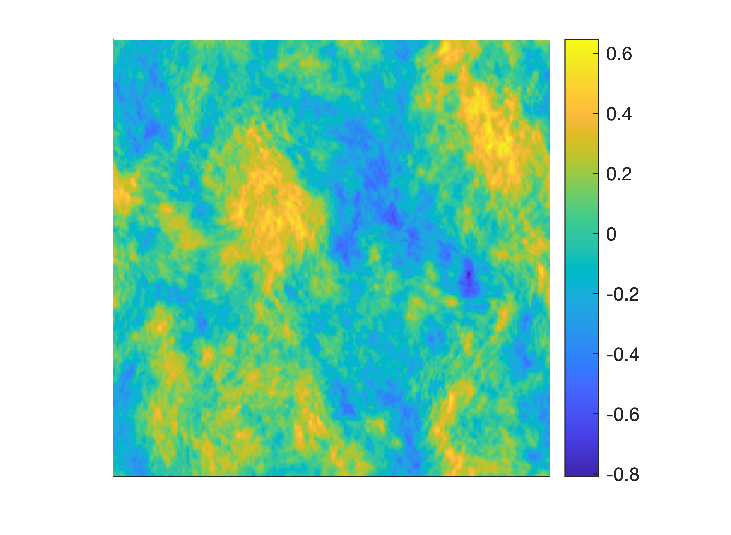}}
    \caption{Instantaneous snapshot of the $y$-component of velocity at an arbitrary $x-y$ plane.}
    \label{fig:Gabormodes_hit_snapshots}
\end{figure*}

\begin{figure}[htp]%[!ht]
\centering
\setlength{\textfloatsep}{10pt plus 1.0pt minus 2.0pt}
\setlength{\floatsep}{6pt plus 1.0pt minus 1.0pt}
\setlength{\intextsep}{6pt plus 1.0pt minus 1.0pt}
{\includegraphics[width=1\linewidth]{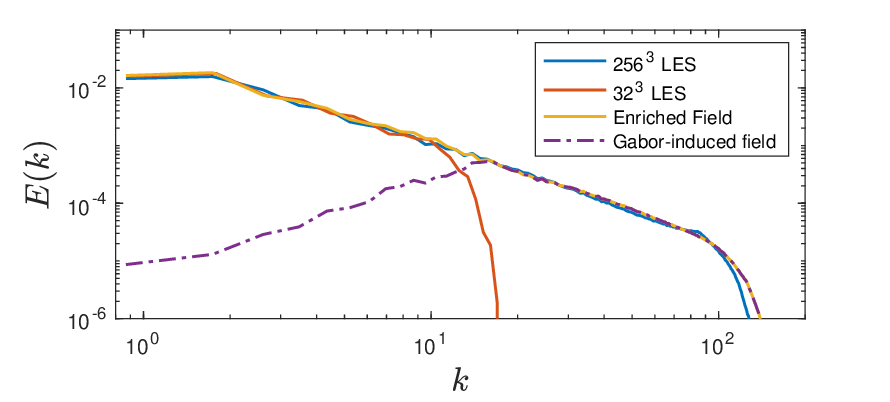}}\\
\caption{Kinetic energy spectrum for forced HIT simulations.}
\label{fig:3D energy spectrum}
\end{figure}

Returning to the setting of HIT, the small scale velocity field reconstruction given by Eq. (\ref{Synthesize field (discrete)}) \color{black} is subject to the RDT straining procedure and then \color{black} superposed with the LES field. A visualization of this superposition is provied in Fig. (\ref{fig:Gabormodes_hit_snapshots}) where contours of the $y$-component of velocity are shown for a $32^3$ LES, Gabor-induced small-scales, and the resulting enriched field. The enrichment procedure accurately captures two-point correlations as seen by the excellent agreement of the energy spectra in Fig. (\ref{fig:3D energy spectrum}). This spectral extrapolation is possible with traditional kinematic simulations (such as Fung et. al \cite{fung1992kinematic}) where a Fourier representation is used given the triple periodicity of the problem in question. However, such a reconstruction of the velocity field is unable to capture interscale energy transfer which is a fundamentally local (in physical space) process as demonstrated in \cite{ghate_lele_2020}. The Gabor enrichment method on the other hand, accurately predicts not only the mean value of interscale energy transfer but its distribution (see \cite{ghate_lele_2020} for a detailed discussion). 

\begin{figure}[htp]%[hbt!]
    \centering
	\subfigure[Temporal autocorrelation of small scales]{\includegraphics[trim={0.75cm 0 0.75cm 0},clip,width=.45\textwidth]{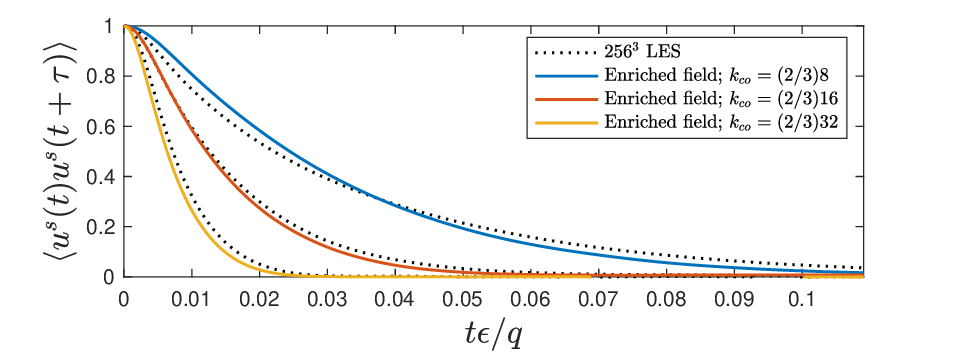}}
	\subfigure[Temporal autocorrelation of full field]{\includegraphics[trim={0.75cm 0 0.75cm 0},clip,width=.45\textwidth]{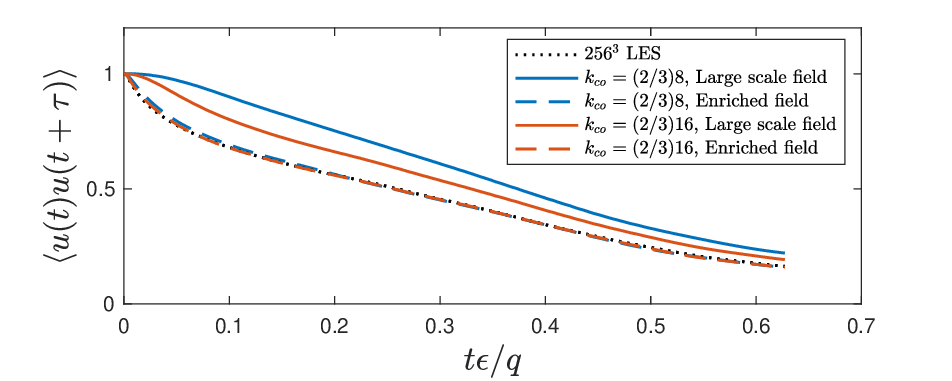}}
	\caption{\textit{a priori} study of the decorrelation of (a) the Gabor induced small scales, and (b) the full enriched field compared to a $256^3$ benchmark LES. A variety of cutoff wavenumbers are considered.}
    \label{fig:Velocity decorrelation}
\end{figure}

Accurate temporal statistics are an additional necessary condition for a satisfactory enrichment method and as such the temporal autocorrelation of velocity is evaluated where the initialized Gabor modes are simply advected by the large scale field. This simplistic dynamic model does not capture the effects of large scale straining of small scales and the nonlinear relaxation due to small scales interacting with each other, both of which are accounted for in the full model, Eq. (\ref{Evolution equations}) (see \cite{ghate_lele_2020} for derivation). However, by reducing to the sweeping model we demonstrate the power of the spatially localized Gabor modes to accurately capture second order temporal statistics as seen in Fig. (\ref{fig:Velocity decorrelation}). 

\begin{equation}\label{Evolution equations}
\begin{aligned}
    \frac{d a_i}{dt} &= \bigg( \frac{2 k_i k_m}{k^2} - \delta_{im} \bigg) a_j \frac{\partial U_m}{\partial x_j} - (\nu + \nu_t(k)) k^2 a_i \\
    \frac{dk_i}{dt} &= -k_j \frac{\partial U_j}{\partial x_i} \\
    \frac{dx_i}{dt} &= U_i
\end{aligned}
\end{equation}

It is worth noting that space-time correlations are difficult to model. The challenge is described in detail by He et. al. \cite{he2017space}. The authors state that the key difficulty in isotropic turbulence is the undetermined link between Eulerian and Lagrangian time correlations. In the Eulerian frame sweeping dominates small-scale decorrelation, while in the Lagrangian frame straining dominates; when modeling these correlations it is a challenge to reconcile the two frames of reference in order to generate a physically consistent flow field. 

One model that has shown to be robust in a variety of flows is the Elliptic Approximation (EA) model described in \cite{he2017space}. In this model, a convection and sweeping velocity must be specified in conjunction with a known one-time, two-point correlation function. The model captures both Taylor's frozen flow and Kraichnan-Tennekes' random sweeping models to predict space-time correlations of the velocity field. However, note that in the present enrichment method no such parameter specification is required; the physical mechanisms associated with decorrelation are directly captured, not modeled, due to the spatial and spectral locality of the Gabor modes which evolve with the large scale field. A further evaluation of space-time correlations in the context of the pressure field is included below.

\section{The small and large scale pressure fields}\label{Pressure}

Accurate prediction of small-scale energetics, as shown in the previous section, is one necessary condition for an effective scale enrichment method. A further validation is analyzing the energy content of the pressure field as a function of scale. In other words, we wish to evaluate the enrichment method's ability to extrapolate the pressure spectrum of a LES.

\subsection{A priori analysis}

\begin{figure}[htp]%[!ht]
\setlength{\textfloatsep}{10pt plus 1.0pt minus 2.0pt}
\setlength{\floatsep}{6pt plus 1.0pt minus 1.0pt}
\setlength{\intextsep}{6pt plus 1.0pt minus 1.0pt}
{\includegraphics[width=1\linewidth]{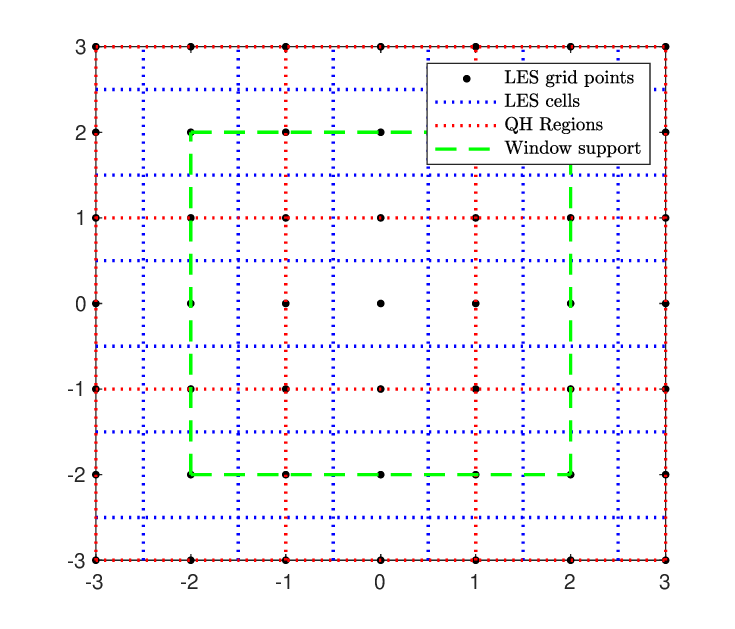}}\\
\caption{Definition of a quasi-homogeneous region for the simulations presented in this paper. The dashed green lines mark the region of influence of a single Gabor mode located at (0,0) due to its window function $f_{\epsilon}$.}
\label{fig:LES subdomain and definition of QH region}
\end{figure}

A $256^3$ LES of forced HIT was run to a statistically stationary state. The simulation imposed zero molecular viscosity so as to imitate the infinite Reynolds number limit and relied exclusively on the subgrid-scale (SGS) model to balance TKE production. We used the Sigma SGS model of Nicoud et. al \cite{nicoud2011using} with model constant, $C_{\sigma} = 0.95$. The flow is forced in spectral space where a specified number, $N$ of Fourier modes with wavenumbers $1 \le |\bm{k}| \le 2$ are randomly chosen each time step. These modes are then subject to an external force

\[
    \hat{f}_i = \frac{\epsilon}{N} \frac{\hat{u}_i}{|\hat{\bm{u}}|^2}
\]

\noindent added to the right-hand side of the Navier Stokes equations thereby prescribing the overall energy dissipation rate. This is the same forcing procedure discussed in \cite{carati1995representation}. Further details of the simulation are provided in \cite{GhateThesis}. An instantaneous flow field was then filtered with a spectrally sharp filter with a cutoff wavenumber equal to $(2/3)16$ thereby giving spectral resolution equivalent to a $32^3$ LES. The filtered velocity field was then enriched with Gabor modes (512 modes per quasi-homogeneous (QH) region) and compared to the original field. Fig. (\ref{fig:LES subdomain and definition of QH region}) shows the definition of a QH region for the simulations considered in this paper. Details on the choice of QH region size and window function support width are given in \cite{GhateThesis}.

The pressure spectrum defined by Eq. (\ref{Pressure Spectrum}) was computed for each case. $P$ corresponds to fluctuating pressure; $\hat{P}$ is the Fourier transform of pressure; $\hat{P}^*$ is the complex conjugate of $\hat{P}$; and the subscript $k$ on the averaging operator denotes an average over wavenumber shells. The enriched field is a superposition of the filtered large-scale field and the Gabor-induced sub-filter field. After superposition, pressure is computed by solving the Poisson equation for pressure exactly in spectral space.

\begin{equation}\label{Pressure Spectrum}
    E_p(k) = \langle \hat{P}^*(\bm{k}) \hat{P}(\bm{k}) \rangle_k
\end{equation}

\begin{figure}[htp]%[!ht]
\setlength{\textfloatsep}{10pt plus 1.0pt minus 2.0pt}
\setlength{\floatsep}{6pt plus 1.0pt minus 1.0pt}
\setlength{\intextsep}{6pt plus 1.0pt minus 1.0pt}
{\includegraphics[width=1\linewidth]{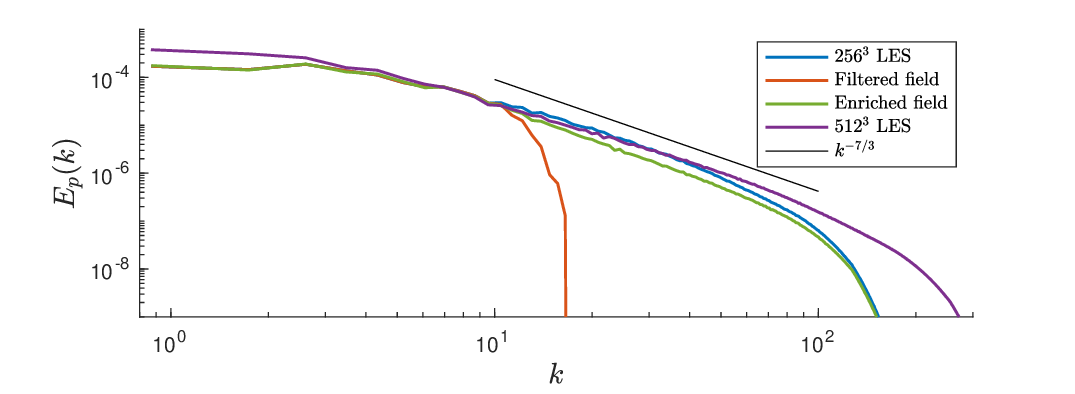}}\\
\caption{Pressure spectra comparison (\textit{a priori} analysis). Note the $512^3$ curve is for comparison only. The \textit{a priori} analysis is in reference to the $256^3$ simulation. The discrepancy in energy content at low wavenumbers in the $512^3$ case is a simple matter of choosing a random snapshot in time for comparison.}
\label{fig:Pressure spectra a priori}
\end{figure}

\noindent Fig. (\ref{fig:Pressure spectra a priori}) shows that the enrichment method extends the bandwidth of scales present by nearly a decade. The large scale field is unable to capture the inertial subrange but the pressure field, enriched with the Gabor modes, extends the bandwidth such that there is an inertial range comparable to the $256^3$ LES in bandwidth, but actually better predicts the $k^{-7/3}$ behavior expected in the infinite Reynolds number limit (see for example \cite{GotohPhysRev2001}, \cite{monin2013statistical}, \cite{batchelor1953theory}, \cite{george1984pressure}). This is demonstrated by including the spectrum from a $512^3$ simulation which follows the $k^{-7/3}$ behavior for a longer range than the $256^3$ LES is able to capture. A deeper investigation of the asymptotic scaling in the form of pre-multiplied spectra are not reported due to the limited scale range for inertial range behavior. While the total energy content is not exact in the enriched field, as shown by the vertical gap in the spectra, the difference in total energy is minimal and the predicted behavior of the inertial range is quite satisfactory.

The ability of the enriched field to predict pressure variance is evaluated as a function of cutoff wavenumber as well. Table (\ref{Pressure stats table}) quantifies these results showing the enrichment method consistently improves the prediction of pressure variance as well as pressure gradient variance. Figure(\ref{fig:Error in statistics}) plots the relative error of the enriched field compared to the baseline $256^3$ LES. Error is defined by Eq. (\ref{error definition}) where $Q$ represents a generic quantity of interest:

\begin{figure}[htp]%[!ht]
\centering
\setlength{\textfloatsep}{10pt plus 1.0pt minus 2.0pt}
\setlength{\floatsep}{6pt plus 1.0pt minus 1.0pt}
\setlength{\intextsep}{6pt plus 1.0pt minus 1.0pt}
{\includegraphics[width=1\linewidth]{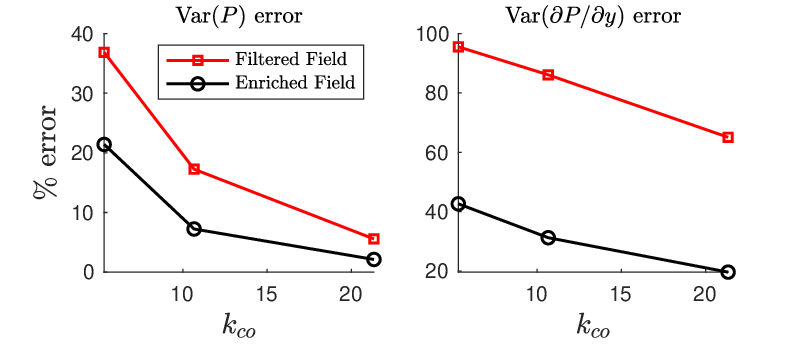}}\\
\caption{Relative error in pressure statistics as a function of cutoff wavenumber}
\label{fig:Error in statistics}
\end{figure}

\begin{equation}\label{error definition}
    \% Error = \frac{\lvert Q_{\text{filtered field}} - Q_{256^3 \text{ LES}}\rvert}{Q_{256^3 \text{ LES}}} \times 100
\end{equation}

\noindent Variance has the typical definition:

\[
    \text{Var}(Q) = \langle Q^2 \rangle_x
\]

\noindent The angle brackets with subscript $x$ indicate a domain-wide spatial average. 

\begin{figure}[htp]%[hbt!]
    \centering
	\subfigure[Log scale]{\includegraphics[trim={0.75cm 0 0.75cm 0},clip,width=.45\textwidth]{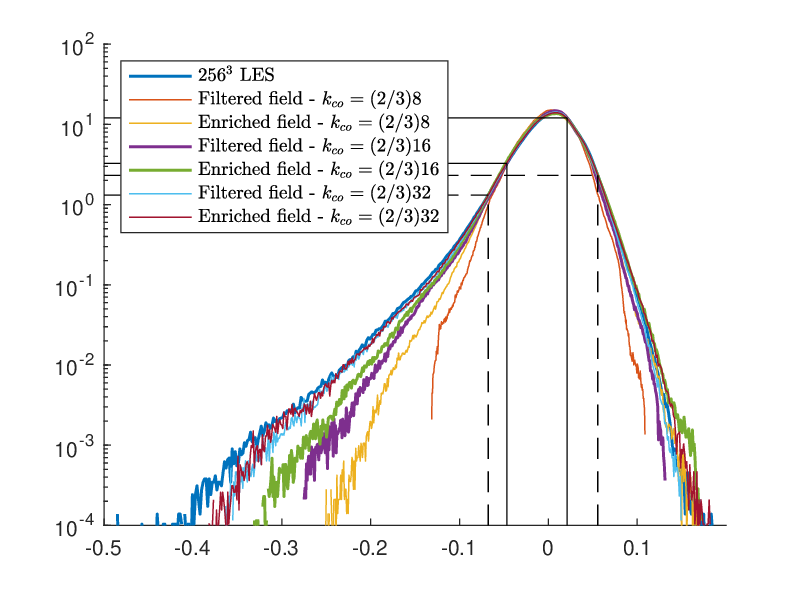}}
	\subfigure[Linear scale]{\includegraphics[trim={0.75cm 0 0.75cm 0},clip,width=.45\textwidth]{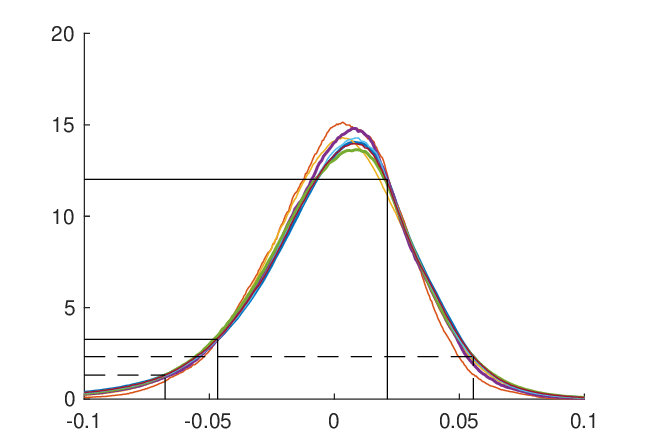}}
    \caption{Probability density function of pressure. The top figure is plotted on a logarithmic vertical axis to highlight the tails of the distribution whereas the bottom figure is displayed on a linear plot to show the comparison near the peak. The solid and dashed black lines mark where the filtered ($k_{co} = (2/3)16$) and enriched PDF, respectively, diverge from the benchmark $256^3$ case.}
    \label{fig:Pressure PDF overlays}
\end{figure}

\begin{figure}[htp]%[hbt!]
    \centering
	\subfigure[Log scale]{\includegraphics[trim={0.75cm 0 0.75cm 0},clip,width=.45\textwidth]{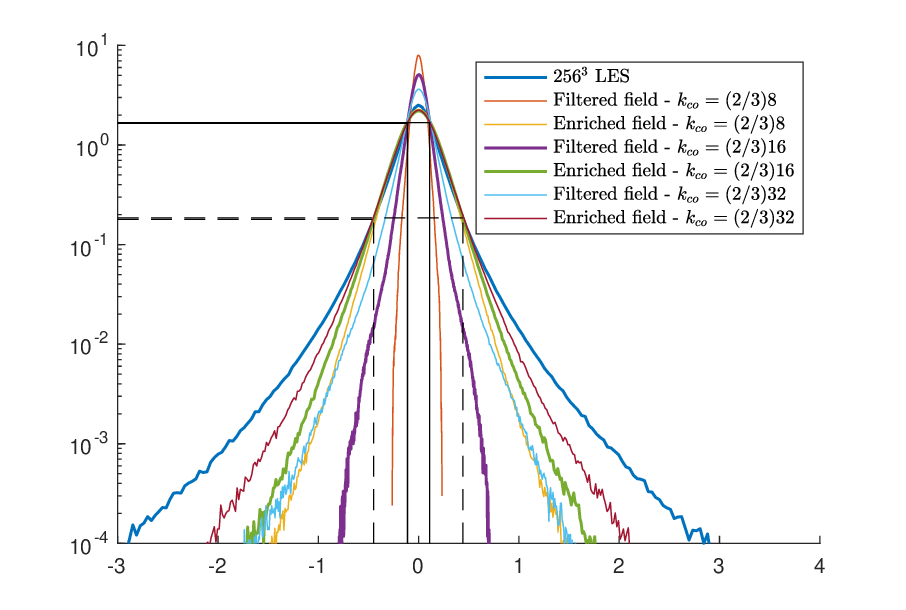}}
	\subfigure[Linear scale]{\includegraphics[trim={0.75cm 0 0.75cm 0},clip,width=.45\textwidth]{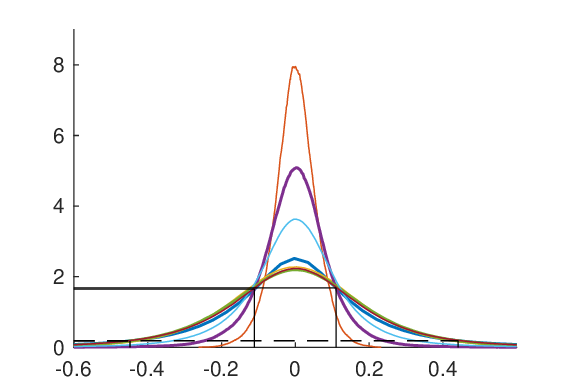}}
    \caption{Probability density function of $y$-derivative of pressure. The solid and dashed black lines correspond to the same markings as Fig. (\ref{fig:Pressure PDF overlays}).}
    \label{fig:dPdy PDF overlays}
\end{figure}

Figures (\ref{fig:Pressure PDF overlays}) and (\ref{fig:dPdy PDF overlays}) are included to show the qualitative agreement in PDF shape for the enriched field. Both sets of PDFs are displayed on logarithmic and linear axes to highlight the comparison of the tails and peaks of the distributions respectively. The pressure PDF is determined predominantly by the large scales hence the filtered field PDFs match the benchmark case reasonably well. However, the enrichment method still improves things quantitatively. This can be seen by considering the variance of pressure in Table (\ref{Pressure stats table}) where the error is reduced by a factor of 1.7, 2.4, and 2.6 for cutoff wavenumbers $(2/3)8$, $(2/3)16$, and $(2/3)32$ respectively. The solid and dashed black lines on the pressure PDF (Fig. (\ref{fig:Pressure PDF overlays})) are to mark where the PDFs of the filtered field ($k_{co}=(2/3)16$) and its enriched counterpart respectively, overlap with the baseline case. By integrating the region between the two we find that events with $27\%$ probability of occurring are represented accurately in the enriched field but are under-represented in the filtered field.

The comparison to the pressure gradient\footnote{The pressure gradient is here represented as only the $y$-derivative of pressure. This being isotropic turbulence the other components are indistinguishable.} PDF is particularly striking as the small-scale contribution dominates the statistics of pressure derivatives. We see significant improvement for all cutoff wavenumbers considered. Lines marking the intersection of the filtered ($k_{co}=(2/3)16$) and enriched cases are included as in the pressure PDF. Here we find that events with $44\%$ probability of occurring are represented accurately in the enriched field as opposed to the filtered field where they are significantly misrepresented or missing altogether. Figure (\ref{fig:Error in statistics}) shows significant improvement in pressure-derivative variance error for all cutoff wavenumbers.

\begin{table}%[h!]
\centering
    \begin{tabular}{|c|c|c|c|c|}
    \hline
       Case  &  Var($P$) & \% Error& Var($\partial P/ \partial y$) & \% Error \\ \hline
        $256^3$ LES (baseline) & 0.0012 & 0 & 0.0710 & 0 \\ \hline
        F, $k_{co} = (2/3)8$ & 0.0008 & 36.85 & 0.0032 & 95.55 \\
        E, $k_{co} = (2/3)8$ & 0.0009 & 21.41 & 0.0407 &  42.66 \\
        \hline
        F, $k_{co} = (2/3)16$ & 0.0010 & 17.26 & 0.0099 & 86.10 \\
        E, $k_{co} = (2/3)16$ & 0.011 & 7.23 & 0.0488 & 31.24 \\
        \hline
        F, $k_{co} = (2/3)32$ & 0.0011 & 5.57 & 0.0248 & 65.07 \\
        E, $k_{co} = (2/3)32$ & 0.0012 & 2.12 & 0.0570 & 19.65 \\
        \hline
    \end{tabular}
    \caption{Pressure statistics. ``F'' and ``E'' denote the \textit{filtered} field and its \textit{enriched} counterpart respectively. 
    }
    \label{Pressure stats table}
\end{table}

\subsubsection{Space-time correlations}

In addition to reliable extrapolation of the pressure spectrum we must also consider the temporal behavior of the pressure field in order to validate the method's ability to produce a physically consistent representation, one that will have practical importance and this necessitates investigation of the space-time autocorrelation of pressure, $R_P(\bm{r},\tau)$ defined as:

\begin{equation}\label{Pressure space-time correlation}
    R_P(\bm{r},\tau) = \langle P(\bm{x},t) P(\bm{x}+\bm{r},t + \tau) \rangle_{x,t}
\end{equation}

\noindent where the subcripts $x$ and $t$ signify spatial and temporal averaging respectively (the spatial averaging is over all three coordinate directions unless otherwise specified). $R_P(\bm{r},\tau)$ can be conveniently expressed in terms of the time-correlations of various \textit{scales} or wavenumbers by taking the spatial Fourier transform of Eq. (\ref{Pressure space-time correlation}) and averaging over wavenumber shells:

\[
\begin{aligned}
    c_P(k,\tau) &= \frac{\langle \hat{R}_P(\bm{k},\tau) \rangle_k}{\langle \hat{R}_P(\bm{k},0) \rangle_k }; \\ \hat{R}_P(\bm{k},\tau) &= 
    \int_{\bm{r} \in \mathbb{R}^3}
    R_P(\bm{r},\tau) e^{-i \bm{k} \cdot \bm{r}} d\bm{r}
\end{aligned}
\]

Here the correlation is expressed in terms of the normalized autocorrelation coefficient. Additionally, both the longitudinal and transverse space-time correlations of the pressure gradient are computed; these are denoted as $f$ and $g$ respectively and are defined as

\begin{equation}\label{long and trans correlations}
\begin{aligned}
    f(r,\tau) &= \frac{\big \langle \nabla P_p (\bm{x},t) \nabla P_p(\bm{x}+\bm{r},t + \tau) \big \rangle_{x,t}}{\big \langle (\nabla P_p)^2 \big \rangle_{x,t}} \\
    g(r,\tau) &= \frac{\big \langle \nabla P_n(\bm{x},t) \nabla P_n(\bm{x}+\bm{r},t + \tau) \big \rangle_{x,t}}{\big \langle (\nabla P_n)^2 \big \rangle_{x,t}}
\end{aligned}
\end{equation}

\noindent where the subscripts $p$ and $n$ denote the pressure gradient components \textit{parallel} and \textit{normal} to the separation vector $\bm{r}$. If we denote each direction with the unit vectors $\bm{p}$ and $\bm{n}$ respectively, these are chosen such that 

\[ 
\begin{aligned} 
\nabla P_p &\equiv \nabla P \cdot \bm{r} \text{ and } \bm{p} \cdot \bm{n} = 0 
\end{aligned}
\]

\begin{figure}[htp]%[!ht]
\centering
\setlength{\textfloatsep}{10pt plus 1.0pt minus 2.0pt}
\setlength{\floatsep}{6pt plus 1.0pt minus 1.0pt}
\setlength{\intextsep}{6pt plus 1.0pt minus 1.0pt}
{\includegraphics[width=1\linewidth]{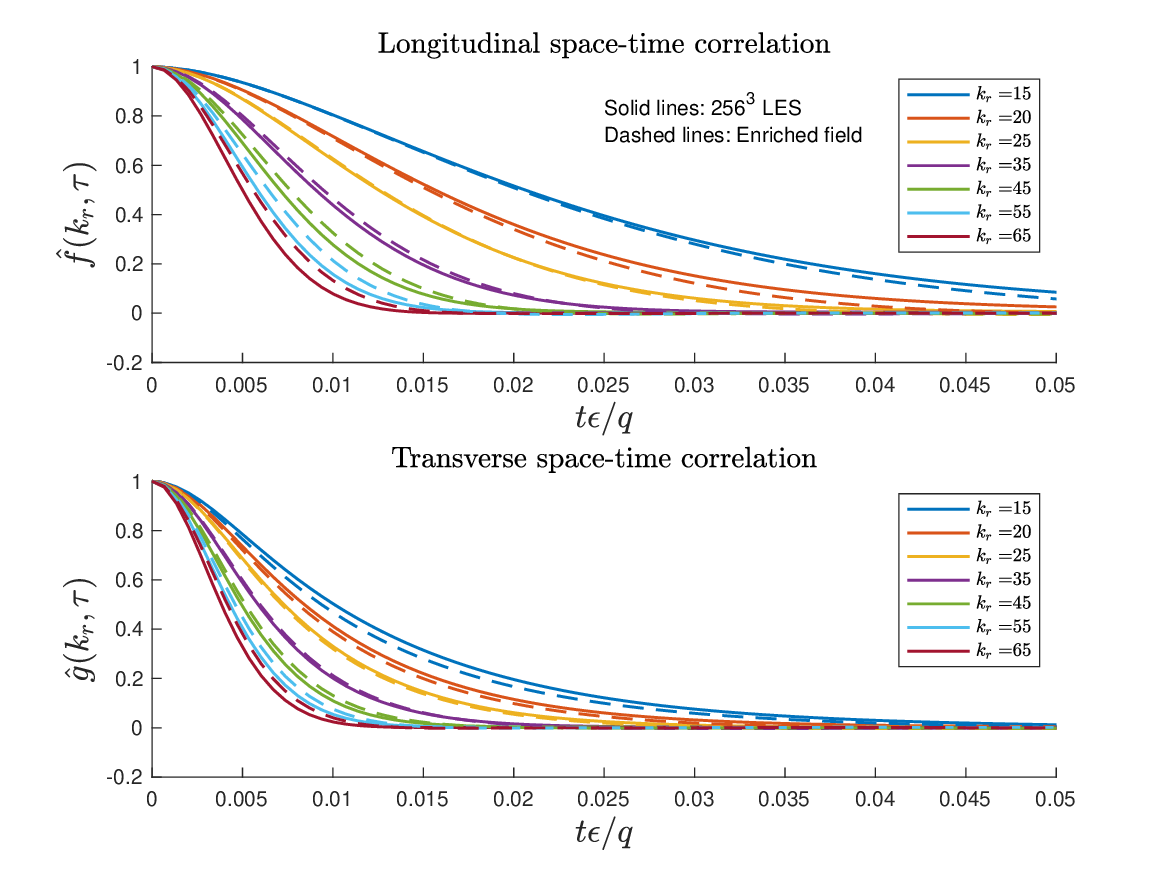}}\\
\caption{Longitudinal and transverse space-time correlation of the pressure gradient (\textit{a priori}).}
\label{fig:Pressure grad space-time apriori}
\end{figure}

\noindent Clearly the choice of $\bm{n}$ is not unique. In practice these correlations were computed as

\[
\begin{aligned}
    \big \langle (\nabla P_1)^2 \big \rangle f(r) &= C_{11}(r,0,0) \\
    \big \langle (\nabla P_2)^2 \big \rangle g(r) &= C_{22}(r,0,0)
\end{aligned}
\]

\noindent where $C_{ij}(\bm{r}) \equiv \big\langle \nabla P_i(\bm{x}) \nabla P_j (\bm{x} + \bm{r}) \big\rangle$. The spatial Fourier transform (in the $\bm{r}$ direction) of the correlations, defined in Eq. (\ref{FT of gradient correlations}), is shown in Fig. (\ref{fig:Pressure grad space-time apriori}) for a variety of wavenumbers. 

\begin{equation}\label{FT of gradient correlations}
\begin{aligned}
    \hat{f}(k_r,\tau) = \int_{r \in \mathbb{R}} f(r,\tau) e^{-i k_r r} dr \\
    \hat{g}(k_r,\tau) = \int_{r \in \mathbb{R}} g(r,\tau) e^{-i k_r r} dr
\end{aligned}
\end{equation}

\begin{figure}[htp]%[!ht]
\centering
\setlength{\textfloatsep}{10pt plus 1.0pt minus 2.0pt}
\setlength{\floatsep}{6pt plus 1.0pt minus 1.0pt}
\setlength{\intextsep}{6pt plus 1.0pt minus 1.0pt}
{\includegraphics[width=1\linewidth]{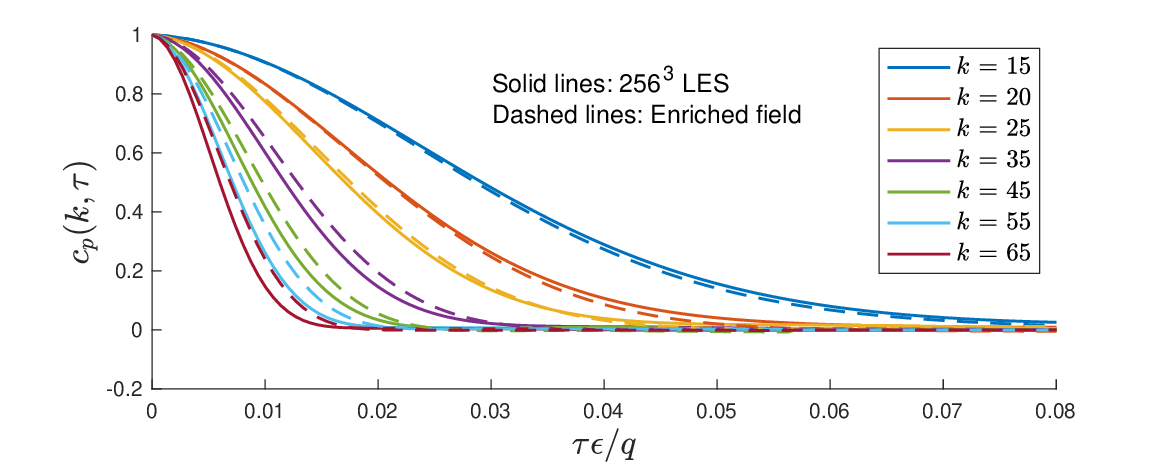}}\\
\caption{Space time correlation of pressure. The $x$-axis is normalized by energy dissipation rate, $\epsilon$, and domain-averaged kinetic energy, $q$. Note that all correlations plotted here are completely absent in the filtered field since the cutoff wavenumber is smaller than 15.}
\label{fig:Pressure space-time correlations}
\end{figure}

The pressure correlations are plotted in Fig. (\ref{fig:Pressure space-time correlations}) and the pressure gradient in Fig. (\ref{fig:Pressure grad space-time apriori}). In computing these correlations the initialized Gabor modes were evolved according to the simple sweeping model discussed in section \ref{Energetics}. In the present case of \textit{a priori} analysis the large scales are represented exactly and so are not compared in the figure, but the information regarding autocorrelations of scales below $k_{co}$ is completely absent and reliant on the enrichment procedure. The plots demonstrate close agreement with the full field. 

\begin{figure}[htp]%[!ht]
\centering
\setlength{\textfloatsep}{10pt plus 1.0pt minus 2.0pt}
\setlength{\floatsep}{6pt plus 1.0pt minus 1.0pt}
\setlength{\intextsep}{6pt plus 1.0pt minus 1.0pt}
{\includegraphics[width=1\linewidth]{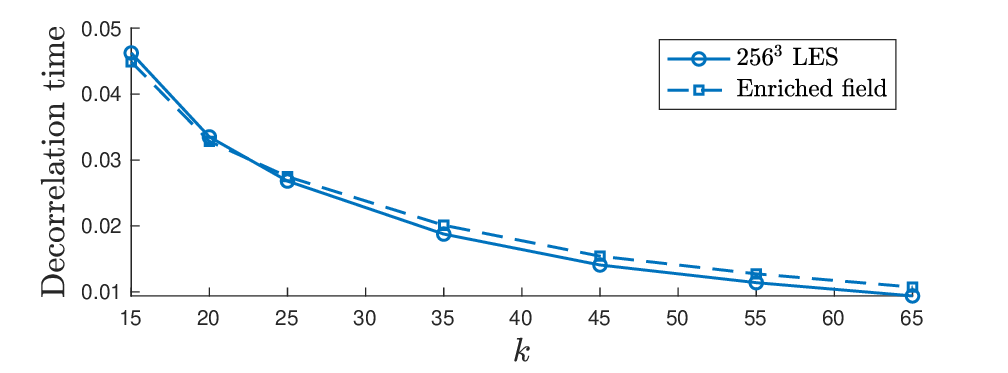}}\\
\caption{Pressure decorrelation time as a function of wavenumber.}
\label{fig:Pressure decorrelation times - a prioir}
\end{figure}

\begin{figure}[htp]%[!ht]
\centering
\setlength{\textfloatsep}{10pt plus 1.0pt minus 2.0pt}
\setlength{\floatsep}{6pt plus 1.0pt minus 1.0pt}
\setlength{\intextsep}{6pt plus 1.0pt minus 1.0pt}
{\includegraphics[width=1\linewidth]{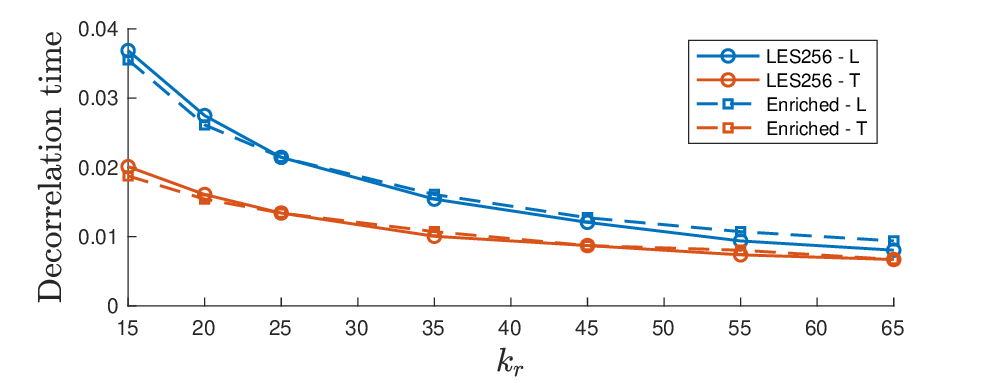}}\\
\caption{Decorrelation times of the pressure gradient longitudinal and transverse correlations.  Legend nomenclature is the same as that in Fig. (\ref{fig:Pressure grad long and trans spectra a priori}).}
\label{fig:Pressure gradient decorrelation times - apriori}
\end{figure}

To quantify this agreement, the decorrelation time, defined as the time for the correlation function to reach 20\% of its initial value, is computed for each wavenumber and plotted in Fig. (\ref{fig:Pressure decorrelation times - a prioir}) for pressure and Fig. (\ref{fig:Pressure gradient decorrelation times - apriori}) for pressure gradient. For both pressure and pressure gradient, we see that the decorrelation time is closely predicted by the enriched field for all wavenumbers with slight under-prediction at small $k_r$ and over-prediction for large $k_r$. The maximum error in decorrelation time is 2.90\%, 4.88\%,  and 6.67\% for pressure, pressure-gradient longitudinal, and pressure-gradient transverse correlations respectively.

\begin{figure}[htp]%[!ht]
\centering
\setlength{\textfloatsep}{10pt plus 1.0pt minus 2.0pt}
\setlength{\floatsep}{6pt plus 1.0pt minus 1.0pt}
\setlength{\intextsep}{6pt plus 1.0pt minus 1.0pt}
{\includegraphics[width=1\linewidth]{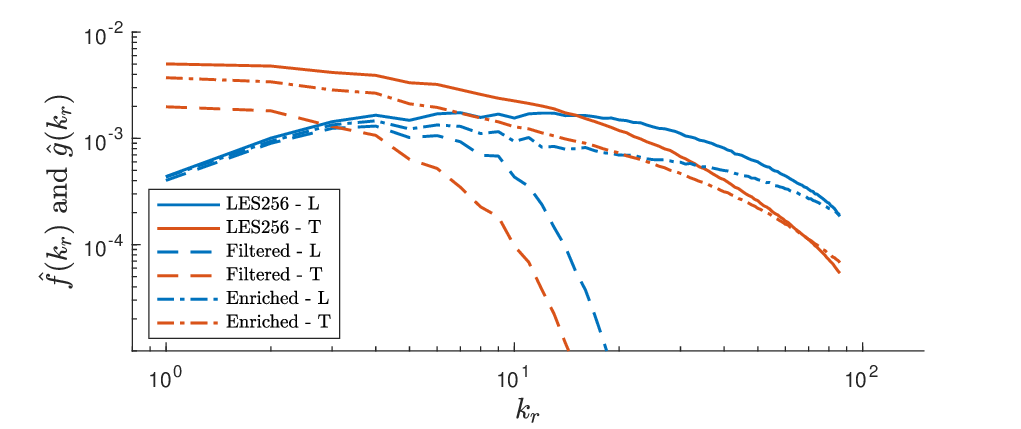}}\\
\caption{Pressure gradient longitudinal and transverse spectra. The nomenclature in the legend is interpreted as follows: ``LES256'' corresonds to the $256^3$ baseline LES. ``Filtered'' is the large scale component of the baseline $256^3$ LES. And ``Enriched'' is the large scale field enriched with Gabor modes. ``L'' and ``T'' correspond to the longitudinal and transverse spectra respectively.}
\label{fig:Pressure grad long and trans spectra a priori}
\end{figure}

In order to better assess the spatial correlations we also plot the one-time longitudinal and transverse spectra for the pressure gradient in Fig. (\ref{fig:Pressure grad long and trans spectra a priori}). We note that for the transverse spectra the enriched field underpredicts the magnitude of the energy at almost all scales except where the curves overlap near the Nyquist wavenumber of the high resolution LES. Additionally, the longitudinal spectra agree reasonably well at the low and high ends of the wavenumber range, but underpredict the energy of intertial-range scales. That said, the improvement of the enriched field over its filtered counterpart is significant.

\subsection{A posteriori analysis}
\begin{figure}[htp]%[!ht]
\setlength{\textfloatsep}{10pt plus 1.0pt minus 2.0pt}
\setlength{\floatsep}{6pt plus 1.0pt minus 1.0pt}
\setlength{\intextsep}{6pt plus 1.0pt minus 1.0pt}
{\includegraphics[width=1\linewidth]{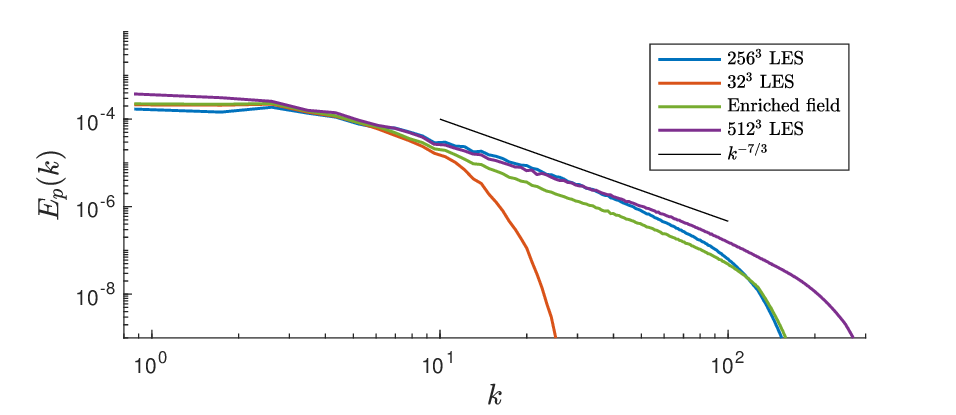}}\\
\caption{Pressure spectra (\textit{a posteriori} analysis)}
\label{fig:Pressure Spectra}
\end{figure}

In contrast to the preceding study here an independent $32^3$ LES was run to a statistically stationary state separate from the $256^3$ benchmark case. A snapshot of the flow field was enriched with Gabor modes and compared against the $256^3$ LES simulation. Results from a separate $512^3$ LES are included with the pressure spectrum plot \color{black} (Fig. [\ref{fig:Pressure Spectra}]) as before to better show the asymptotic $k^{-7/3}$ behavior in the inertial subrange\color{black}. The pressure spectrum for the $256^3$ LES, $32^3$ enriched LES (enriched to $256^3$), and a $512^3$ LES are plotted in Fig. (\ref{fig:Pressure Spectra}).

\begin{figure}[htp]%[hbt!]
    \centering
	\subfigure[Log scale]{\includegraphics[trim={0.75cm 0 0.75cm 0},clip,width=.45\textwidth]{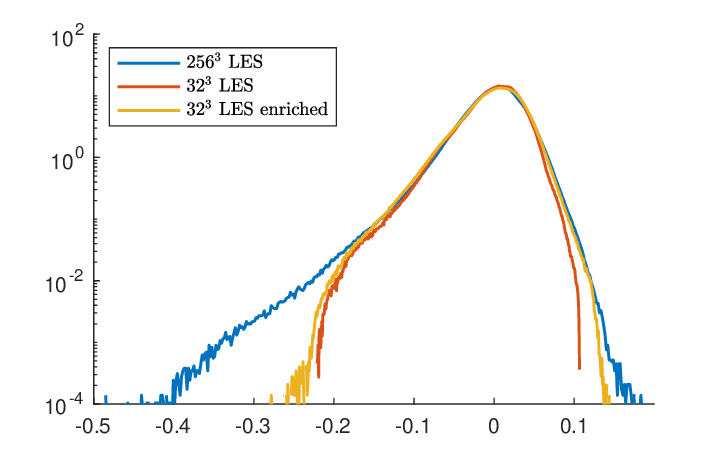}}
	\subfigure[Linear scale]{\includegraphics[trim={0.75cm 0 0.75cm 0},clip,width=.45\textwidth]{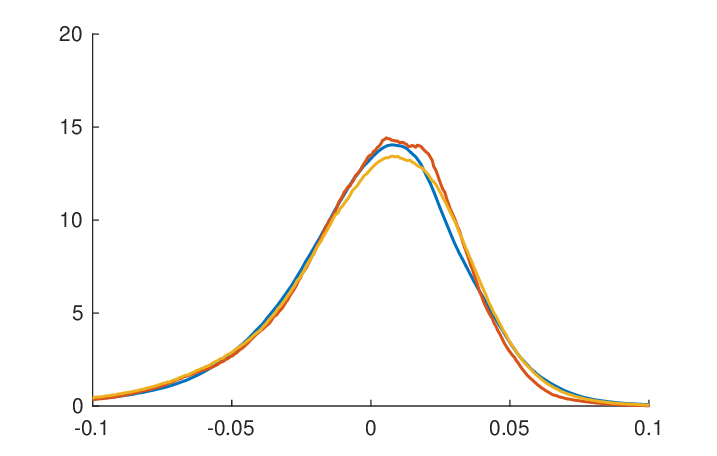}}
    \caption{Probability density functions for pressure (\textit{a posteriori} analysis). The top figure is plotted on a logarithmic vertical axis to highlight the tails of the distribution whereas the bottom figure is displayed on a linear plot to show the comparison near the peak.}
    \label{fig:Pressure PDFs a posteriori}
\end{figure}

\begin{figure}[htp]%[hbt!]
    \centering
	\subfigure[Log scale]{\includegraphics[trim={0.75cm 0 0.75cm 0},clip,width=.45\textwidth]{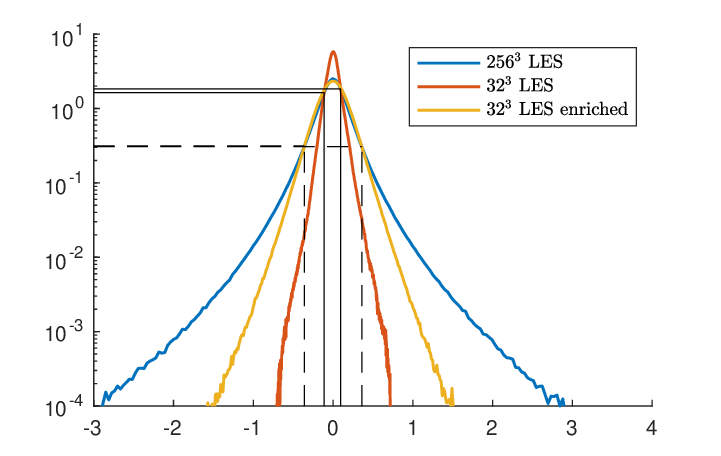}}
	\subfigure[Linear scale]{\includegraphics[trim={0.75cm 0 0.75cm 0},clip,width=.45\textwidth]{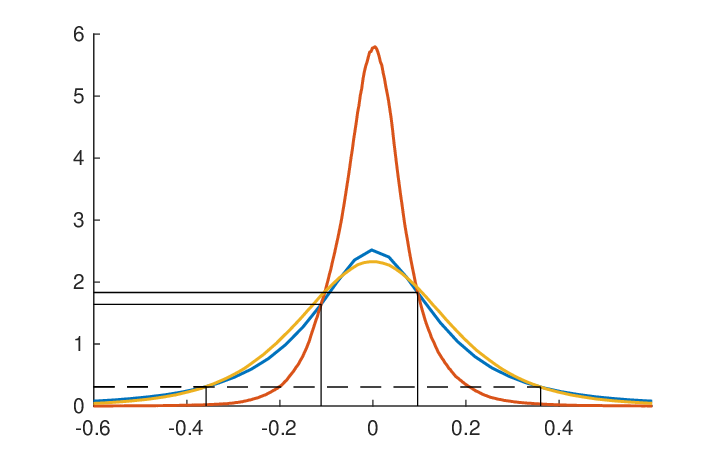}}
    \caption{Probability density functions for $y$-derivative of pressure (\textit{a posteriori} analysis). The solid and dashed black lines mark where the $32^3$ LES and enriched PDF, respectively, diverge from the benchmark $256^3$ case.}
    \label{fig:dPdy PDFs a posteriori}
\end{figure}

We have also included the analysis of pressure and pressure gradient variance. Table (\ref{Pressure stats table 2}) summarizes the results which demonstrates a similar improvement in the variance of both pressure and pressure gradient as seen previously in the \textit{a priori} case. For the present \textit{a posteriori} analysis error is reduced by a factor of 98.6 and 2.1 for pressure and pressure derivative variance respectively\footnote{It should be noted that the extremely large improvement in pressure variance is a bit misleading since these are instantaneous quantities and change from time-step to time-step. That said, the pressure is well represented by the LES and it is the pressure derivative that benefits the most from the enrichment procedure. This is evident by visually comparing the PDFs in Figures (\ref{fig:Pressure PDFs a posteriori}) \& (\ref{fig:dPdy PDFs a posteriori})}. This is a remarkable improvement given there is no knowledge of the baseline large scale field. Additionally, the black lines on the pressure derivative PDF indicate that events with 43\% probability of occurring are missing or significantly under-represented in the $32^3$ LES but are now captured by the enriched field. 

\begin{table}
\centering
    \begin{tabular}{|c|c|c|c|c|}
    \hline
       Case  &  Var($P$) & Error(\%) & Var($\partial P/ \partial y$) & Error(\%) \\ \hline
       $256^3$ LES (baseline) & 0.0012 & 0 & 0.0710 & 0 \\ \hline
        F, $k_{co} = (2/3)8$ & 0.0008 & 36.85 & 0.0032 & 95.55 \\
        E, $k_{co} = (2/3)8$ & 0.0009 & 21.41 & 0.0407 &  42.66 \\
        \hline
        F, $k_{co} = (2/3)16$ & 0.0010 & 17.26 & 0.0099 & 86.10 \\
        E, $k_{co} = (2/3)16$ & 0.0011 & 7.23 & 0.0488 & 31.24 \\
        \hline
        F, $k_{co} = (2/3)32$ & 0.0011 & 5.57 & 0.0248 & 65.07 \\
        E, $k_{co} = (2/3)32$ & 0.0012 & 2.12 & 0.0570 & 19.65 \\
        \hline
        $32^3$ LES & 0.0010 & 12.86 & 0.0084 & 88.22 \\ \hline
        Enriched $32^3$ LES & 0.0012 & 0.13 & 0.0411 & 42.04\\
        \hline
    \end{tabular}
    \caption{Pressure statistics. The \emph{a priori} values are included for comparison.}
    \label{Pressure stats table 2}
\end{table}

\begin{figure}[htp]%[!ht]
\setlength{\textfloatsep}{10pt plus 1.0pt minus 2.0pt}
\setlength{\floatsep}{6pt plus 1.0pt minus 1.0pt}
\setlength{\intextsep}{6pt plus 1.0pt minus 1.0pt}
{\includegraphics[width=1\linewidth]{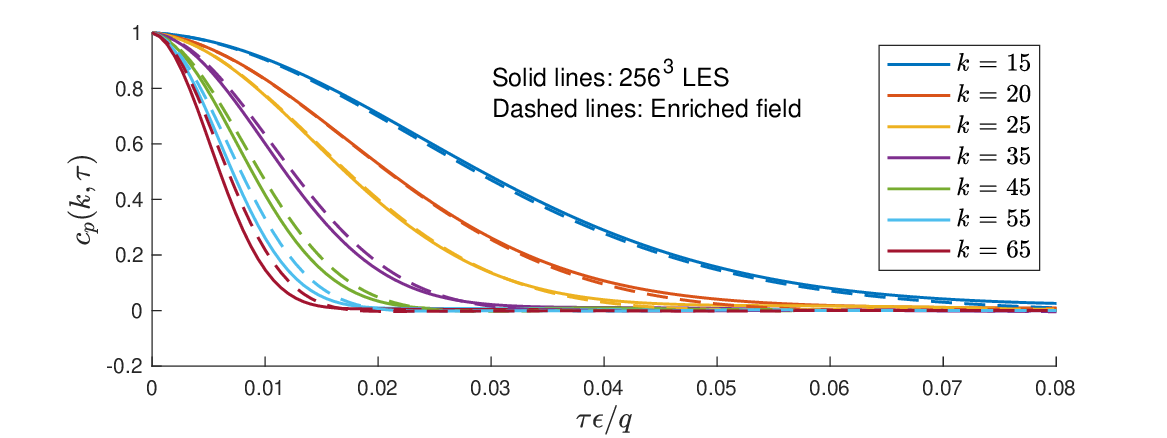}}\\
\caption{Pressure space-time correlations (\textit{a posteriori} analysis). Solid lines are computed from an independent $256^3$ LES and the dashed lines correspond to an enriched $32^3$ LES.}
\label{fig:Pressure space-time correlation (a posteriori)}
\end{figure}

\begin{figure}[htp]%[!ht]
\centering
\setlength{\textfloatsep}{10pt plus 1.0pt minus 2.0pt}
\setlength{\floatsep}{6pt plus 1.0pt minus 1.0pt}
\setlength{\intextsep}{6pt plus 1.0pt minus 1.0pt}
{\includegraphics[width=1\linewidth]{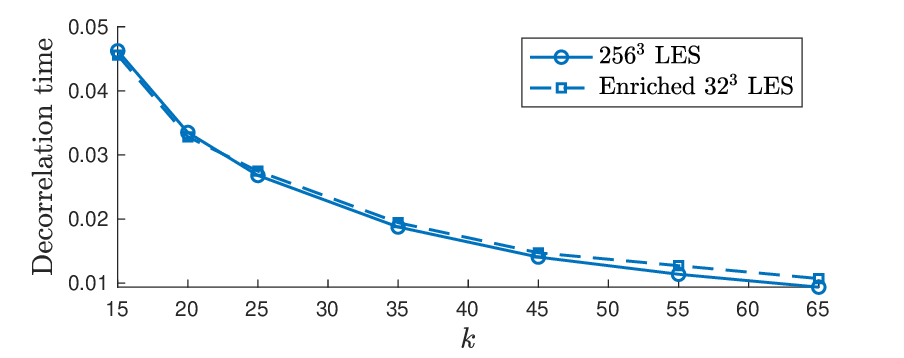}}\\
\caption{Pressure decorrelation time as a function of wavenumber. Decorrelation here is defined as the time when the correlation coefficient reaches 0.2.}
\label{fig:Pressure decorrelation times - a posteriori}
\end{figure}

\begin{figure}[htp]%[!ht]
\centering
\setlength{\textfloatsep}{10pt plus 1.0pt minus 2.0pt}
\setlength{\floatsep}{6pt plus 1.0pt minus 1.0pt}
\setlength{\intextsep}{6pt plus 1.0pt minus 1.0pt}
{\includegraphics[width=1\linewidth]{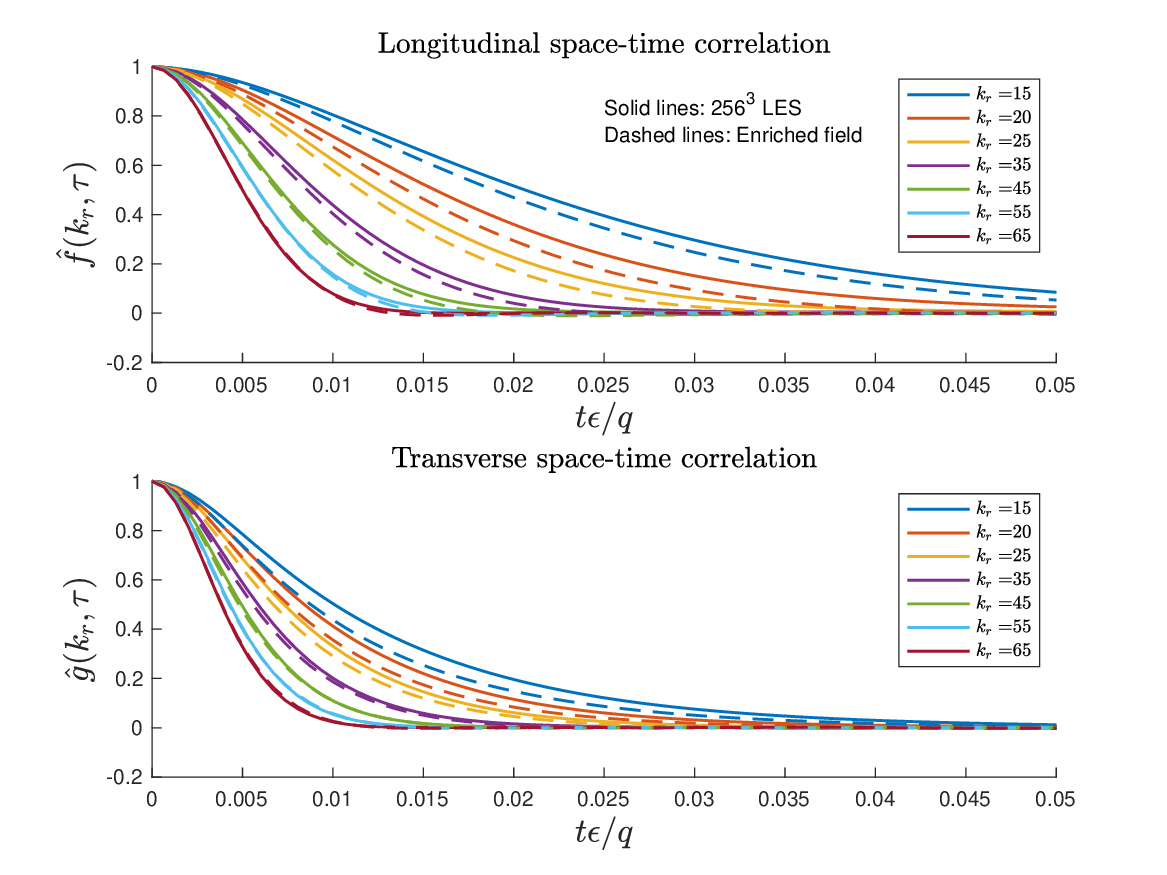}}\\
\caption{Longitudinal and transverse space-time correlation of the pressure gradient (\textit{a posteriori} analysis).}
\label{fig:Pressure grad space-time a posteriori}
\end{figure}

\begin{figure}[htp]%[!ht]
\centering
\setlength{\textfloatsep}{10pt plus 1.0pt minus 2.0pt}
\setlength{\floatsep}{6pt plus 1.0pt minus 1.0pt}
\setlength{\intextsep}{6pt plus 1.0pt minus 1.0pt}
{\includegraphics[width=1\linewidth]{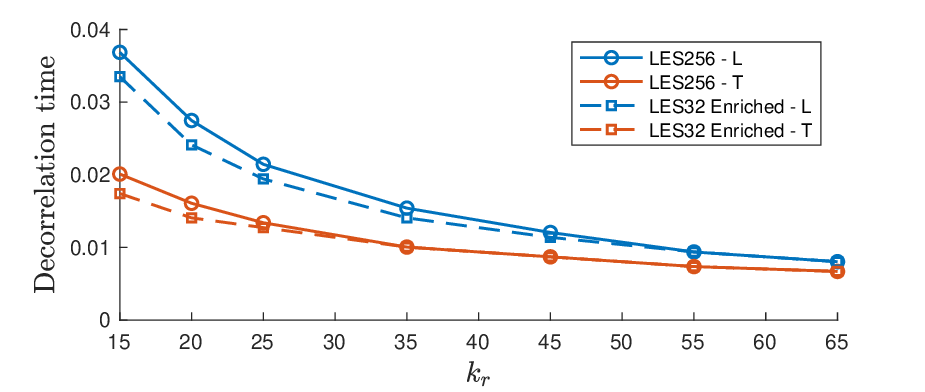}}\\
\caption{Decorrelation times of the pressure gradient longitudinal and transverse correlations. The nomenclature in the legend is the same as Fig. (\ref{fig:Pressure grad long and trans spectra a priori}) with the addition of ``LES32'' corresponding to the $32^3$ LES.}
\label{fig:Pressure grad space-time decorrelation times - a posteriori}
\end{figure}

The space-time correlations are investigated in Figs. (\ref{fig:Pressure space-time correlation (a posteriori)}) and (\ref{fig:Pressure grad space-time a posteriori}) where \color{black}we see the enrichment method is able to closely predict subgrid scale temporal behavior despite the fact the large scales are not ``exact'' (i.e. filtered high-resolution LES)\color{black}. We did not include the large scales in this plot simply because any discrepancy would be the result of the SGS model which has no knowledge of the enriched scales. The decorrelation times are plotted in Figs. (\ref{fig:Pressure decorrelation times - a posteriori}) and (\ref{fig:Pressure grad space-time decorrelation times - a posteriori}). The maximum error in decorrelation time relative to the $256^3$ benchmark LES is 14.29\%, 9.09\%, and 13.33\% for the pressure, pressure-gradient longitudinal, and pressure-gradient transverse correlations respectively.

\begin{figure}[htp]%[!ht]
\centering
\setlength{\textfloatsep}{10pt plus 1.0pt minus 2.0pt}
\setlength{\floatsep}{6pt plus 1.0pt minus 1.0pt}
\setlength{\intextsep}{6pt plus 1.0pt minus 1.0pt}
{\includegraphics[width=1\linewidth]{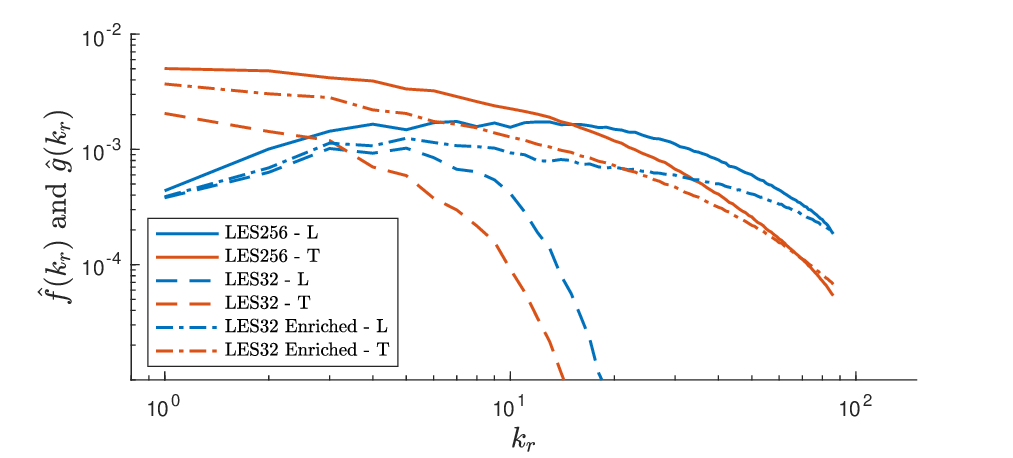}}\\
\caption{Pressure gradient longitudinal and transverse spectra. The nomenclature in the legend is the same as Fig. (\ref{fig:Pressure grad long and trans spectra a priori}) with the addition of ``LES32'' corresponding to the $32^3$ LES.}
\label{fig:Pressure grad long and trans spectra a posteriori}
\end{figure}

And finally, the pressure gradient longitudinal and transverse spectra are plotted in Fig. (\ref{fig:Pressure grad long and trans spectra a posteriori}) showing similar results to the \textit{a priori} case. Again we note the significant improvement over the original $32^3$ LES.  

\section{Decaying HIT}
\begin{figure}[!ht]
\centering
\setlength{\textfloatsep}{10pt plus 1.0pt minus 2.0pt}
\setlength{\floatsep}{6pt plus 1.0pt minus 1.0pt}
\setlength{\intextsep}{6pt plus 1.0pt minus 1.0pt}
{\includegraphics[width=1\linewidth]{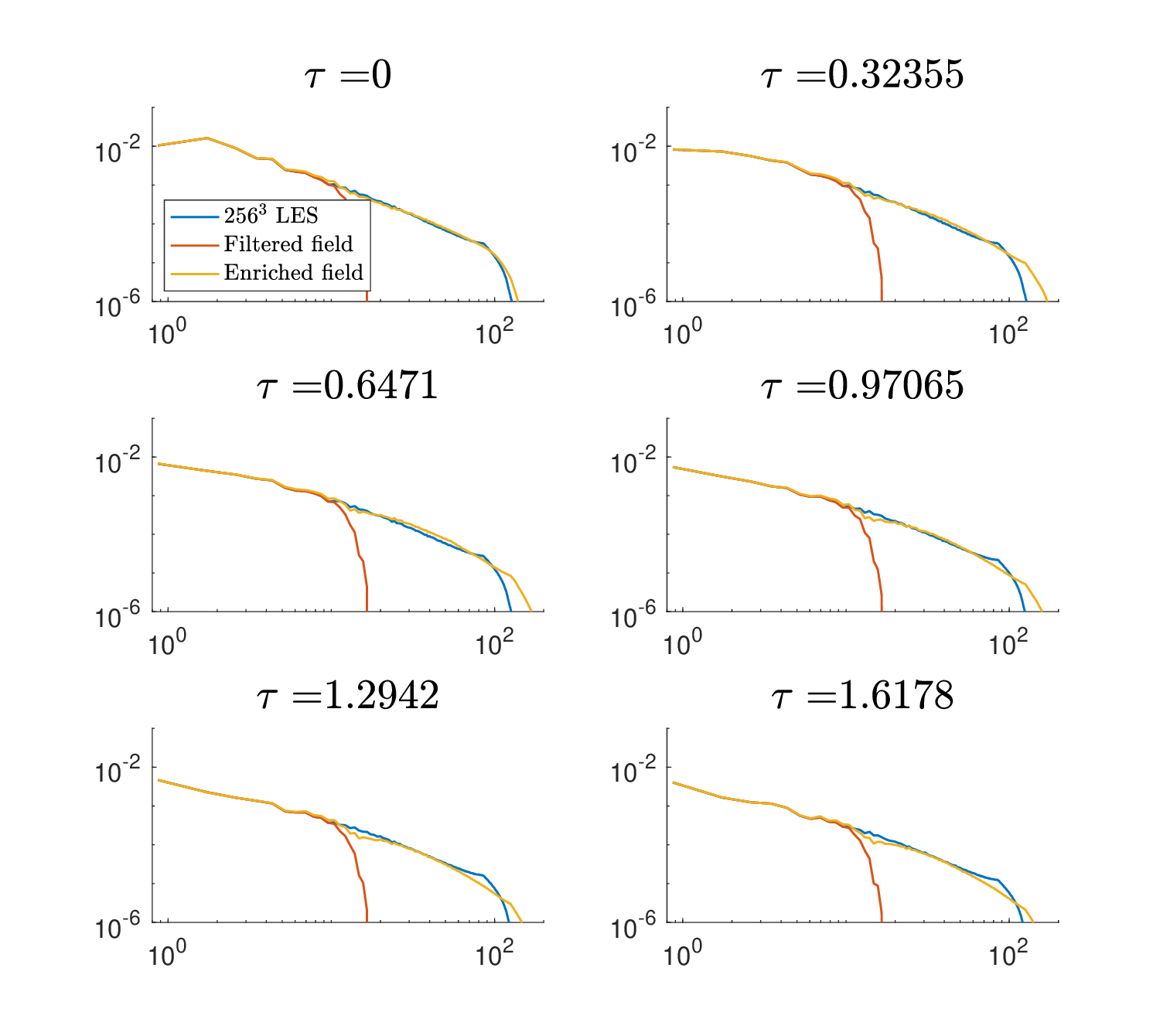}}\\
\caption{Evolution of the energy spectrum in decaying HIT.}
\label{fig:ESpect Decay14}
\end{figure}

The inviscid forced HIT case analyzed above is further integrated after turning the artificial forcing off. The field is filtered using a spectrally sharp filter with cutoff wavenumber, $k_{co} = (2/3)16$. The filtered initial condition is enriched with Gabor modes which then evolve according to Eqs. (\ref{Evolution equations}). Figure (\ref{fig:ESpect Decay14}) shows the energy spectra at various snapshots during the integration. The enrichment method accurately extrapolates the spectral content quite well. However, we note a slight dip of energy near the cutoff wavenumber beginning around $\tau=0.75495$ and the high wavenumber range appears to decay at a slightly faster rate for late times. These issues are currently being investigated with plans to use a dynamic procedure to predict the eddy viscosity coefficient; the current implementation uses a constant value. 

\begin{figure}[!ht]
\centering
\setlength{\textfloatsep}{10pt plus 1.0pt minus 2.0pt}
\setlength{\floatsep}{6pt plus 1.0pt minus 1.0pt}
\setlength{\intextsep}{6pt plus 1.0pt minus 1.0pt}
{\includegraphics[width=1\linewidth]{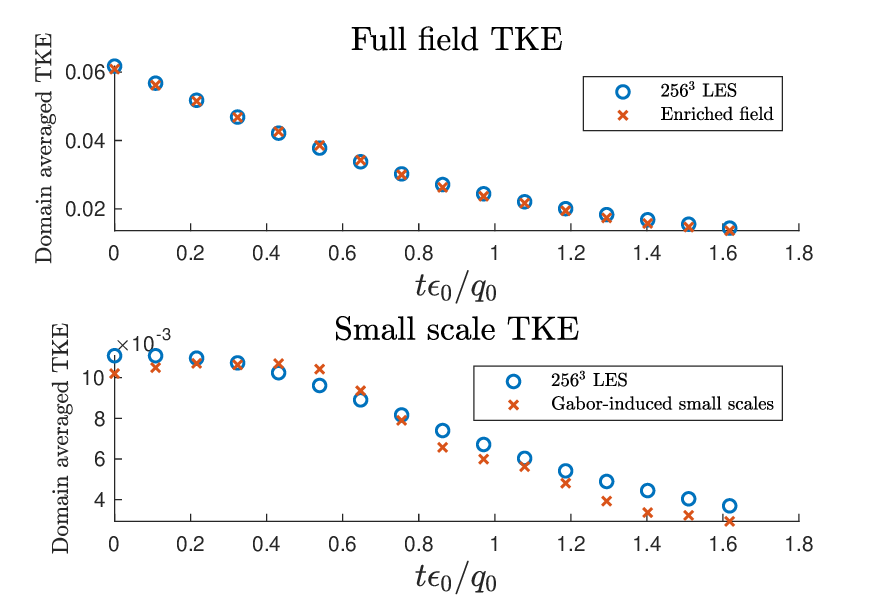}}\\
\caption{Decay of domain-averaged TKE. The top plot shows the kinetic energy for the full field whereas the bottom figure isolates the contribution from the small scales. $\epsilon_0$ and $q_0$ correspond to the kinetic energy dissipation rate and kinetic energy of the large scales at $\tau=0$.}
\label{fig:TKE Decay14}
\end{figure}

Figure (\ref{fig:TKE Decay14}) shows the decay of domain averaged TKE, both for the full field and for the small scales. The decay trend for the small scales is matched by the enrichment procedure but there is noticeable difference between point-wise values. A dynamic model for the spectral eddy viscosity is expected to reduce these differences.

\begin{figure}[!ht]
\centering
\setlength{\textfloatsep}{10pt plus 1.0pt minus 2.0pt}
\setlength{\floatsep}{6pt plus 1.0pt minus 1.0pt}
\setlength{\intextsep}{6pt plus 1.0pt minus 1.0pt}
{\includegraphics[width=1\linewidth]{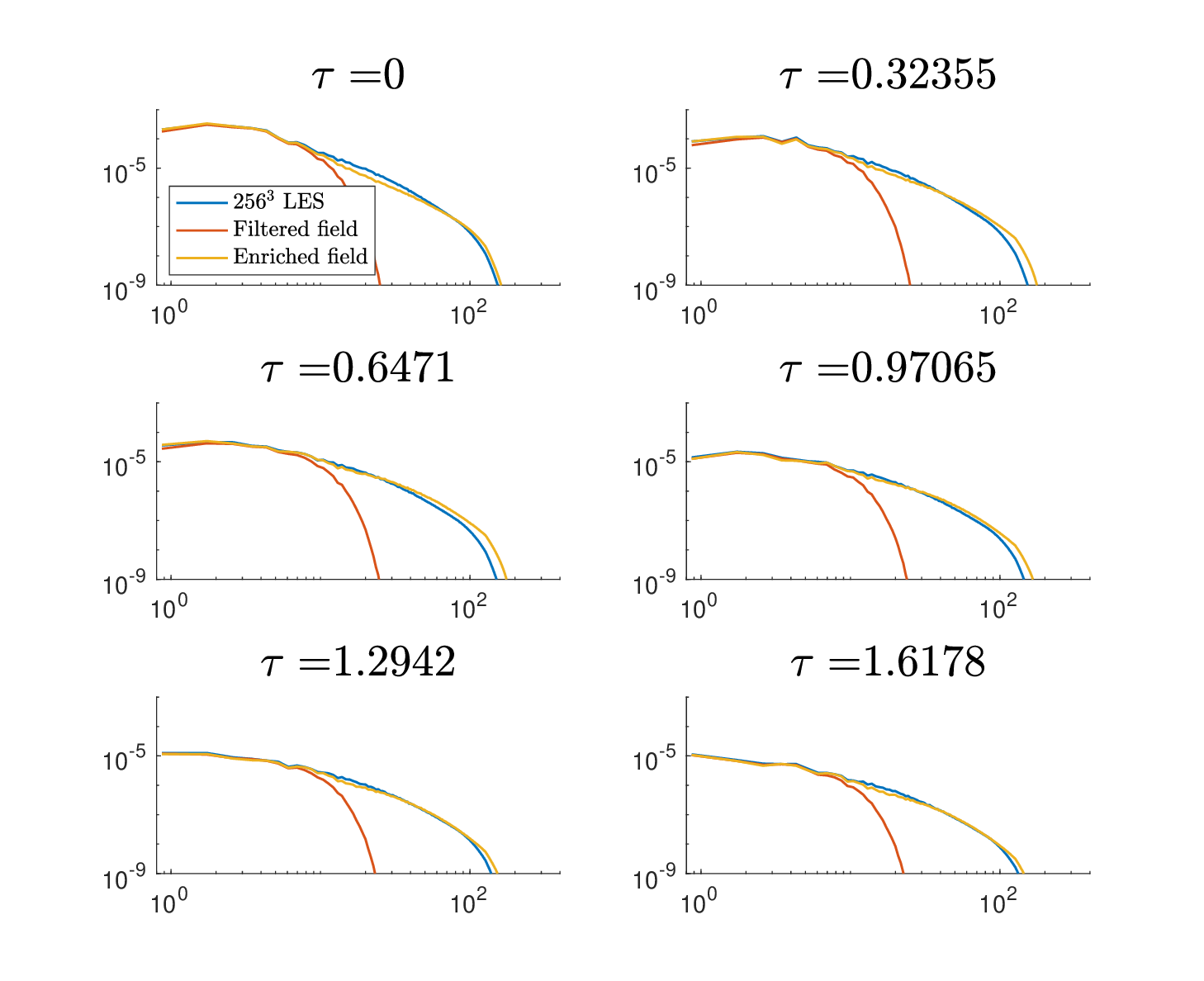}}\\
\caption{Evolution of the pressure spectrum in decaying HIT.}
\label{fig:PSpect Decay14}

\end{figure}

\begin{figure}[!ht]
\centering
\setlength{\textfloatsep}{10pt plus 1.0pt minus 2.0pt}
\setlength{\floatsep}{6pt plus 1.0pt minus 1.0pt}
\setlength{\intextsep}{6pt plus 1.0pt minus 1.0pt}
{\includegraphics[width=1\linewidth]{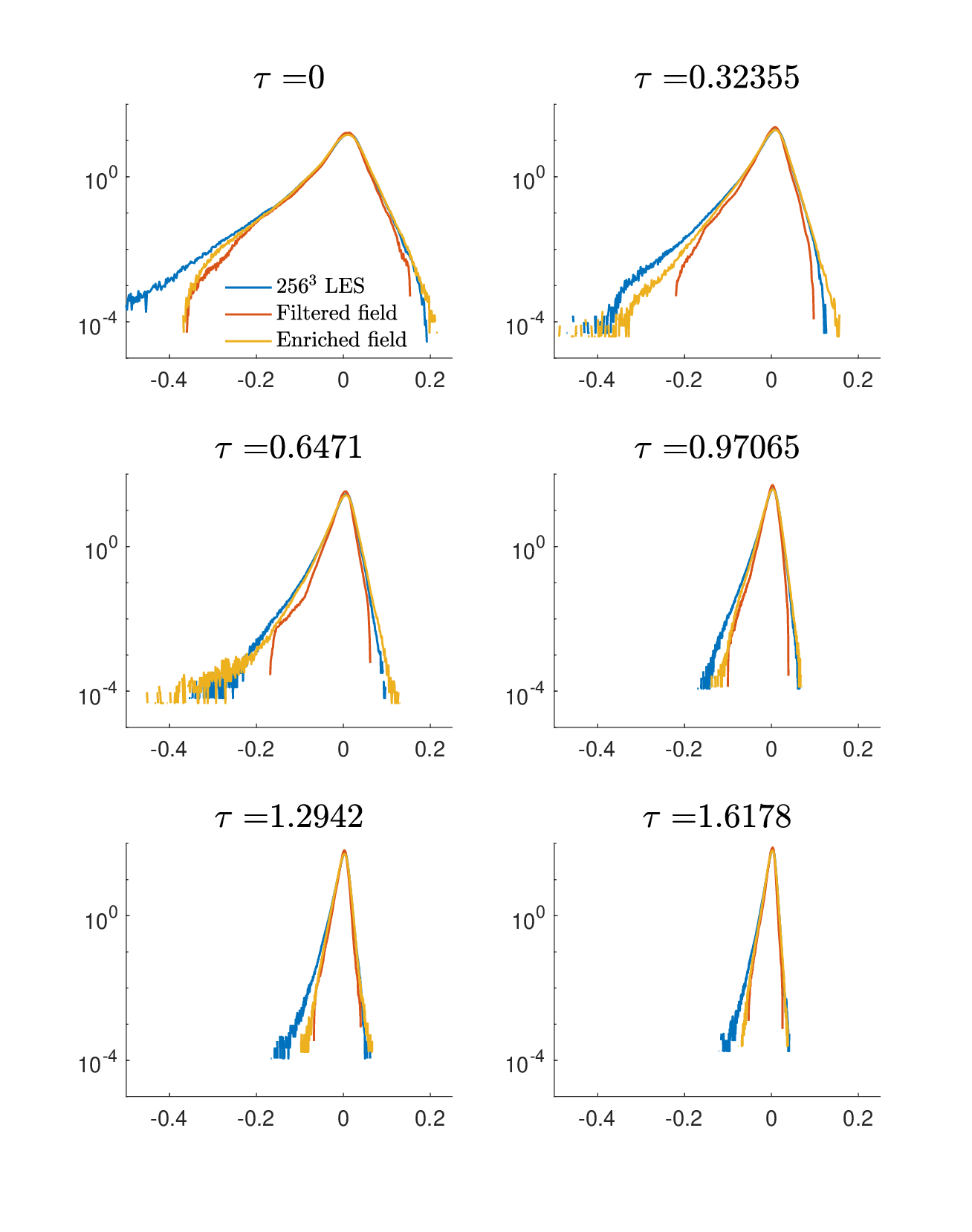}}\\
\caption{Pressure PDFs (log scale)}
\label{fig:Pressure PDF Decay14 - log scale}
\end{figure}

\begin{figure}[!ht]
\centering
\setlength{\textfloatsep}{10pt plus 1.0pt minus 2.0pt}
\setlength{\floatsep}{6pt plus 1.0pt minus 1.0pt}
\setlength{\intextsep}{6pt plus 1.0pt minus 1.0pt}
{\includegraphics[width=1\linewidth]{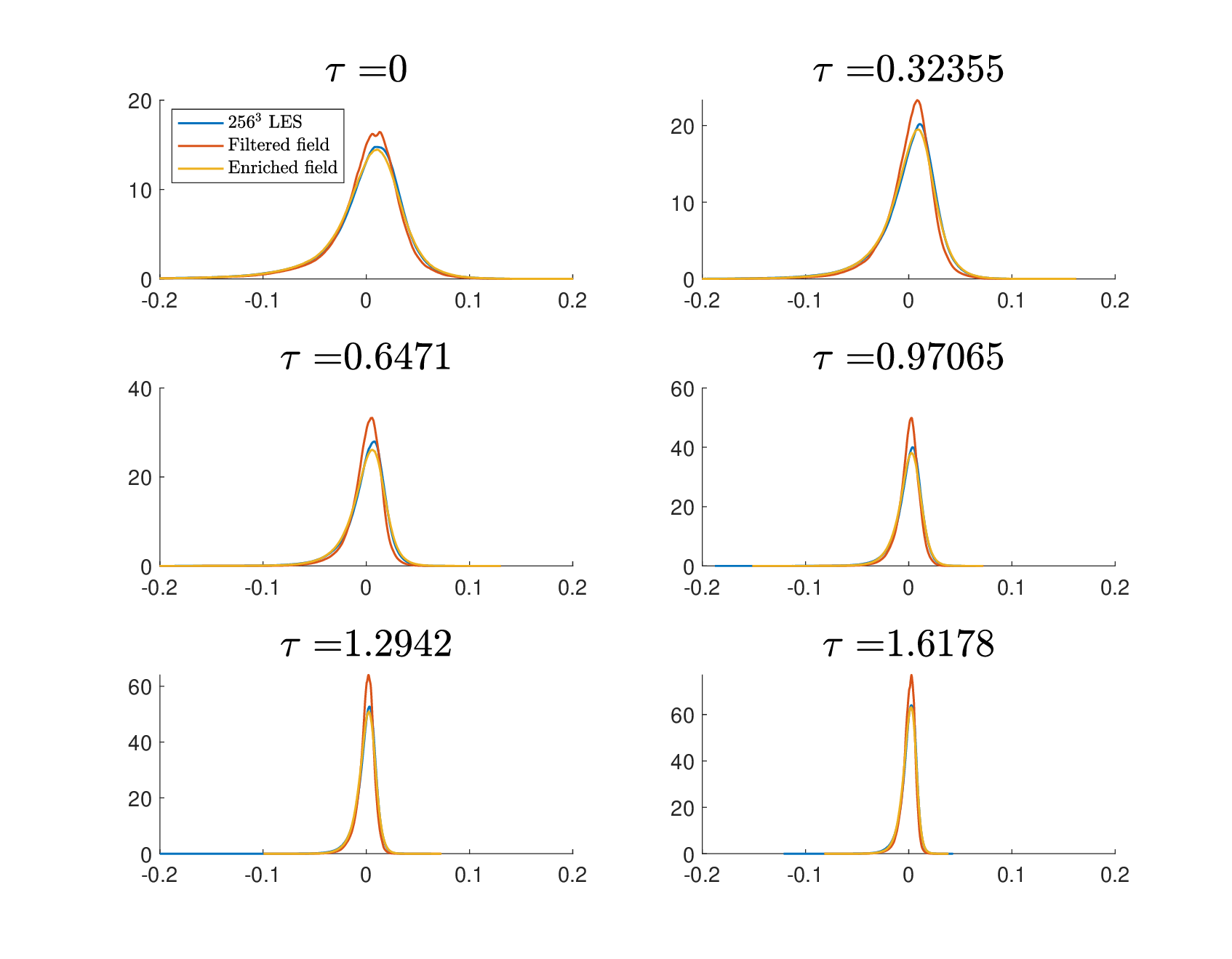}}\\
\caption{Pressure PDFs (linear scale)}
\label{fig:Pressure PDF Decay14}
\end{figure}

\begin{figure}[!ht]
\centering
\setlength{\textfloatsep}{10pt plus 1.0pt minus 2.0pt}
\setlength{\floatsep}{6pt plus 1.0pt minus 1.0pt}
\setlength{\intextsep}{6pt plus 1.0pt minus 1.0pt}
{\includegraphics[width=1\linewidth]{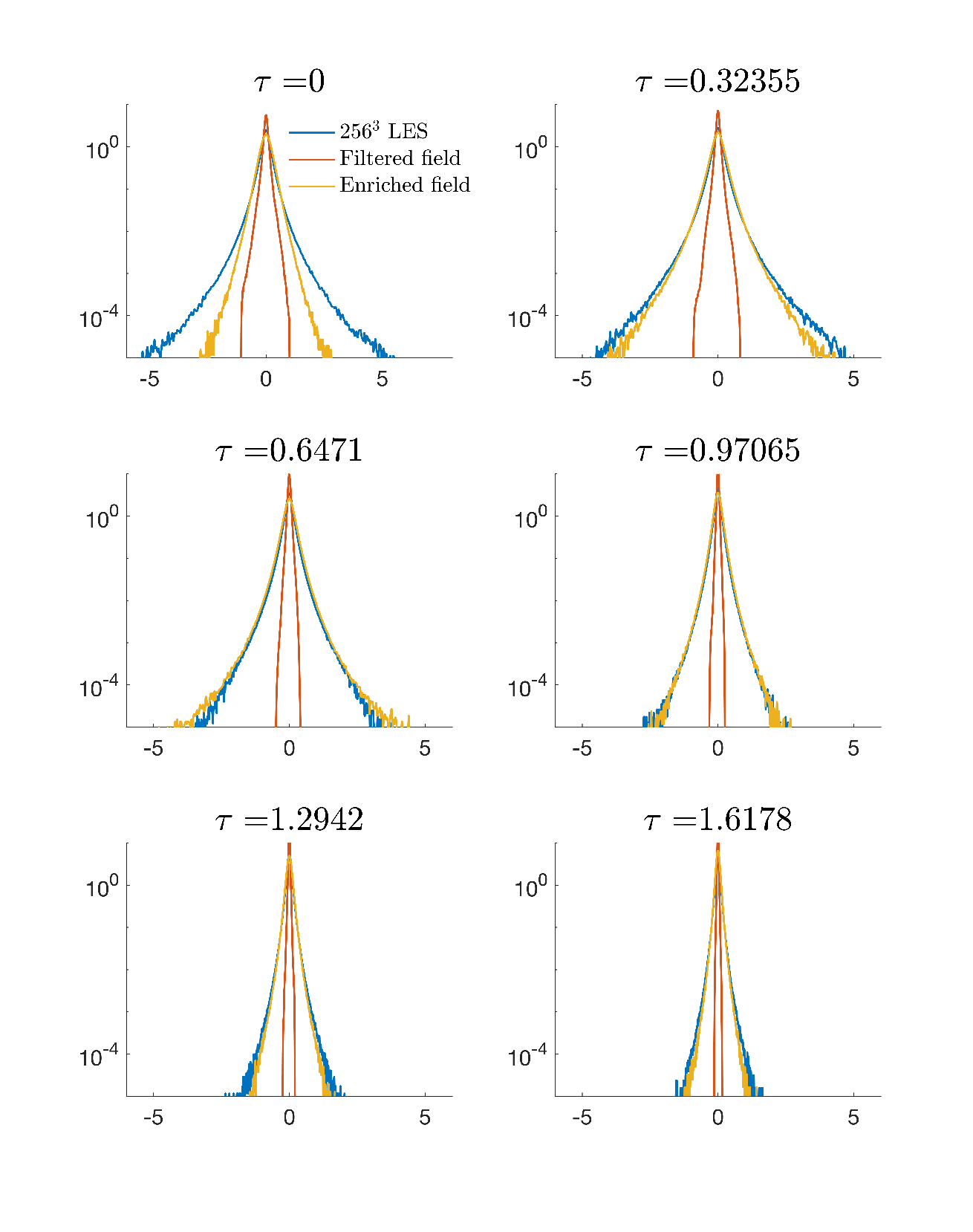}}\\
\caption{$dP/dy$ PDFs (log scale)}
\label{fig:dPdy PDF Decay14 - log scale}
\end{figure}

\begin{figure}[!ht]
\centering
\setlength{\textfloatsep}{10pt plus 1.0pt minus 2.0pt}
\setlength{\floatsep}{6pt plus 1.0pt minus 1.0pt}
\setlength{\intextsep}{6pt plus 1.0pt minus 1.0pt}
{\includegraphics[width=1\linewidth]{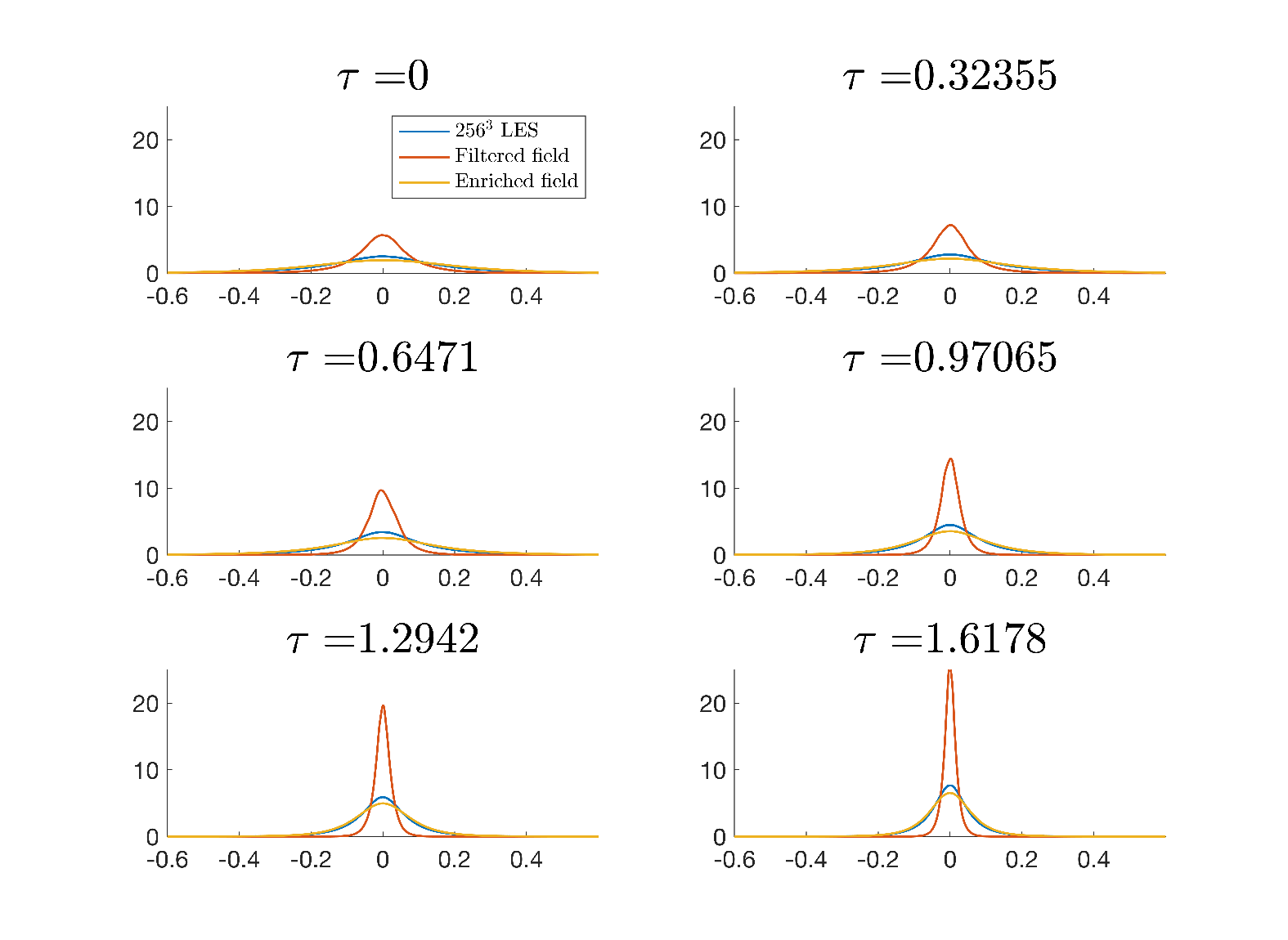}}\\
\caption{$dP/dy$ PDFs (linear scale)}
\label{fig:dPdy PDF Decay14}
\end{figure}

In terms of pressure statistics, Fig. (\ref{fig:PSpect Decay14}) demonstrates the ability of the enriched pressure field to match the decay of the benchmark LES rather well. In fact, compared to the energy spectrum, the pressure content of the enriched field more closely matches the benchmark for all wavenumbers. This is similarly seen in the evolution of both the pressure and pressure derivative PDFs in Figs. (\ref{fig:Pressure PDF Decay14 - log scale})-(\ref{fig:dPdy PDF Decay14}). Particularly striking is how the enriched field appears to relax towards the benchmark (and quite quickly) as evidenced by the top-right plot in Fig. (\ref{fig:dPdy PDF Decay14 - log scale}); the enriched field and benchmark PDFs are nearly on top of each other. There are discrepancies noticeable in the peaks on the linear-scale plots (Fig. [\ref{fig:dPdy PDF Decay14}]), but the performance of the enrichment method at later times is impressive.


\section{Conclusion}

It has been shown that the method described in \cite{GhateThesis}, \cite{GhateJFM2017}, and \cite{ghate_lele_2020} to enrich the LES-resolved velocity field not only predicts the energetics of the subgrid scales but also their contribution to the pressure field. The Gabor-induced field, when superposed with the LES data, does a remarkable job in recovering the high-wavenumber content of the pressure spectrum. This was shown \textit{a priori} where the large scale field is supplied by a filtered high-fidelity LES as well as \textit{a posteriori} where an independent $32^3$ LES is enriched and compared directly to a $256^3$ benchmark LES. In addition to reconstruction of the pressure spectrum, the probability density functions of pressure and pressure derivatives were seen to improve significantly for the enriched simulations. 

We also demonstrated the ability of the enriched field to accurately predict space-time correlations of pressure. The utility of the Gabor modes is that they are physically local and dynamically coupled to the large-scale field thereby the dominant physical mechanism responsible for decorrelation, large scales sweeping small scales, is directly captured. This circumvents the requirement to estimate a sweeping velocity/time scale to predict space-time correlations which is necessary using the random sweeping model of Kraichnan or the EA model of He et. al \cite{he2017space}.

Predicting second order statistics of near-wall pressure fluctuations in wall-bounded flows is of particular interest where pressure fluctuations are responsible for structural vibrations and noise-generation. The ability of WMLES to predict near-wall pressure fluctuations has been investigated by Park \& Moin \cite{park2016space} where they found WMLES capable of predicting pressure variance relatively accurately so long as there is sufficient streamwise and spanwise grid resolution. However, the pressure spectra reported in that paper illustrate that WMLES is incapable of predicting high wavenumber content of the near-wall pressure field. The results shown here from the Gabor-mode enrichment model suggest it is a viable method to enrich the high wavenumber content of near-wall pressure fluctuations. Furthermore, WMLES is unable to capture near-wall shear stress fluctuations (see \cite{park2016space} and \cite{bose2018wall}). This provides a further motivation for enrichment of subgrid scales and another metric with which to evaluate the method in future work.

\begin{acknowledgments}
We wish to acknowledge funding support for this research from the National Science Foundation via grant No. NSF-CBET-1803378.
\end{acknowledgments}

\appendix
\section{Finite Reynolds Number HIT}\label{Appendix - Finite Re DNS}

The simulations considered above impose zero molecular viscosity, relying exclusively on the SGS model to dissipate energy thereby making the dissipative scales grid-resolution-dependent. To eliminate this model dependence from the baseline comparison case we ran a direct numerical simulation (DNS) of a finite Reynolds number flow on a $512^3$ mesh using the same $(2\pi)^3$ domain and forcing procedure as the LES considered in the preceding sections. The simulation achieved $Re_{\lambda} = 221.8$. where  $Re_{\lambda} = u_{rms} \lambda /\nu$ and $\lambda$ is the transverse Taylor microscale (see Eq. (6.57) in \cite{pope2001turbulent}).

\begin{figure}[htp]%[!ht]
    \centering
    \setlength{\textfloatsep}{10pt plus 1.0pt minus 2.0pt}
    \setlength{\floatsep}{6pt plus 1.0pt minus 1.0pt}
    \setlength{\intextsep}{6pt plus 1.0pt minus 1.0pt}
    {\includegraphics[width=1\linewidth]{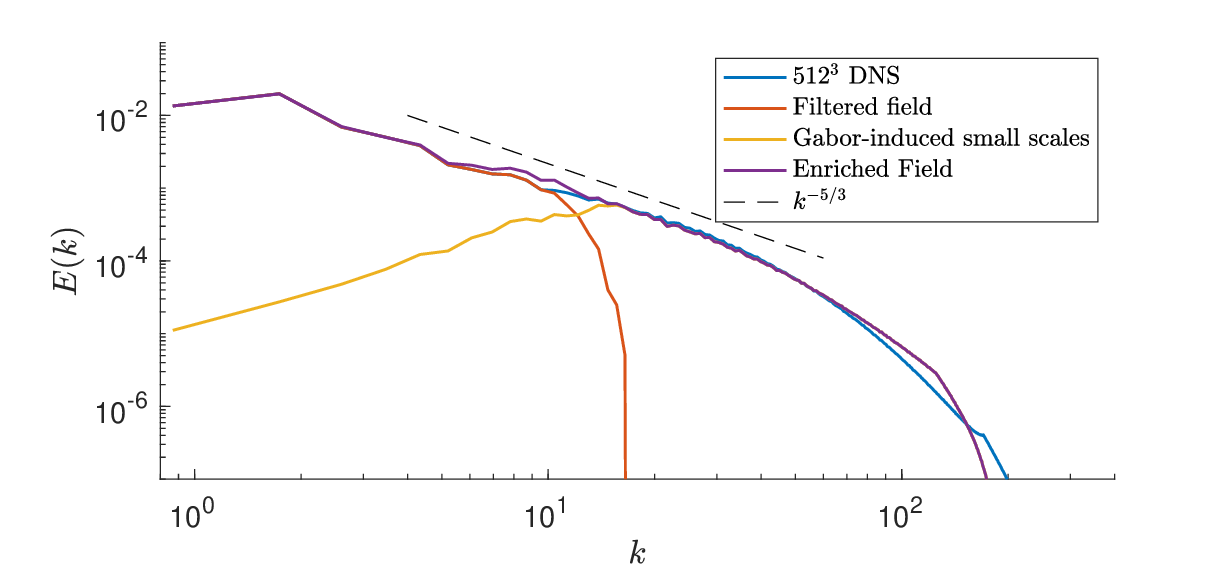}}\\
    \caption{Initial energy spectrum for $512^3$ DNS simulation and it's enriched counterpart}
    \label{fig: DNS512 ESpect}
\end{figure}

The simulation was run for ten eddy-turnover times once a stationary state was achieved before enriching. The initial condition was generated using the same number of modes as that used in the inviscid scenario considered previously which results in the energy spectra seen in Fig. (\ref{fig: DNS512 ESpect}). The subgrid pressure field represented by the Gabor modes generates statistics similar to the inviscid simulation considered above and so our comments on the individual figures will be brief\color{black}; no significant differences are seen in enrichment performance between the inviscid LES and the viscous DNS\color{black}.

\begin{figure}[htp]%[!ht]
    \setlength{\textfloatsep}{10pt plus 1.0pt minus 2.0pt}
    \setlength{\floatsep}{6pt plus 1.0pt minus 1.0pt}
    \setlength{\intextsep}{6pt plus 1.0pt minus 1.0pt}
    {\includegraphics[width=1\linewidth]{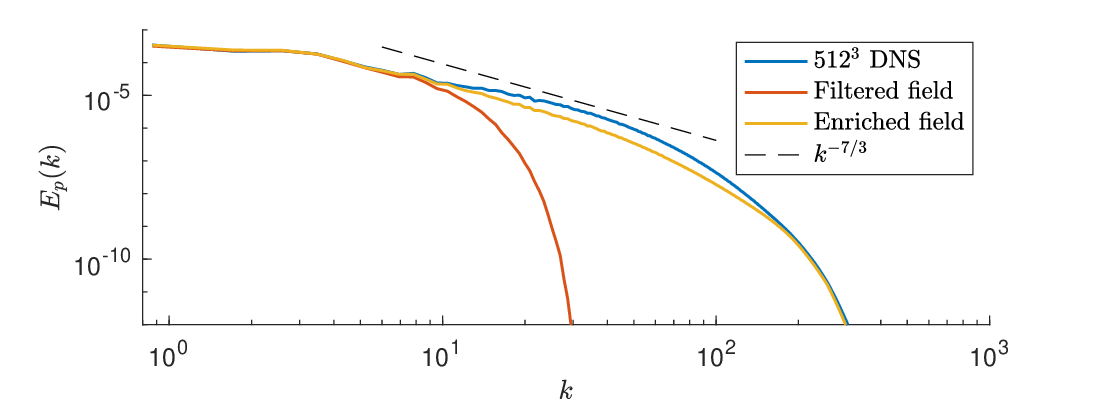}}\\
    \caption{Pressure spectra for finite Re DNS.}
    \label{fig:Pressure spectra DNS512 a priori}
\end{figure}

\begin{figure}[htp]%[hbt!]
    \centering
	\subfigure[Log scale]{\includegraphics[trim={0.75cm 0 0.75cm 0},clip,width=.45\textwidth]{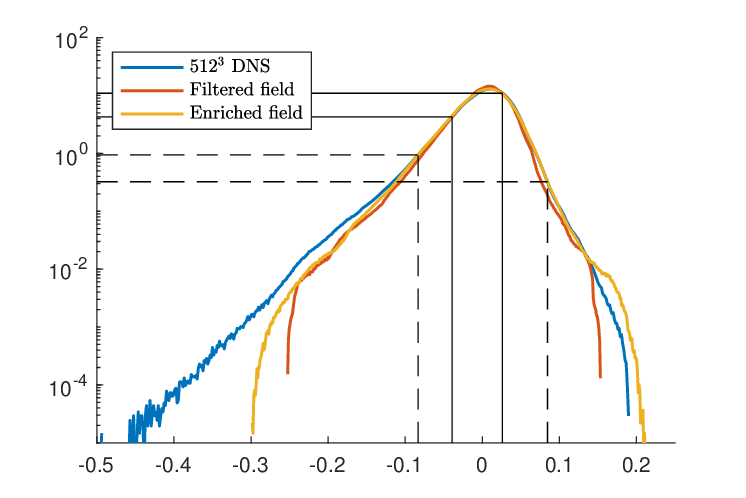}}
	\subfigure[Linear scale]{\includegraphics[trim={0.75cm 0 0.75cm 0},clip,width=.45\textwidth]{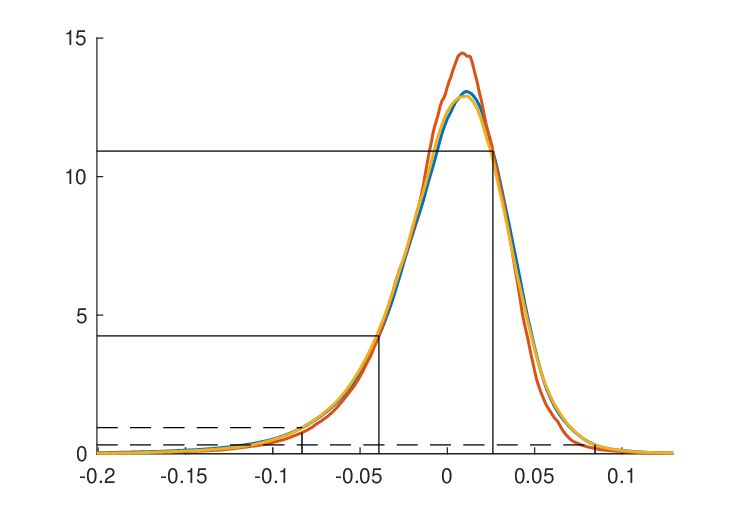}}
    \caption{Probability density functions for pressure. See previous PDF plots for description of solid/dashed black lines.}
    \label{fig:P PDFs DNS512 a priori}
\end{figure}

\begin{figure}[htp]%[hbt!]
    \centering
	\subfigure[Log scale]{\includegraphics[trim={0.75cm 0 0.75cm 0},clip,width=.45\textwidth]{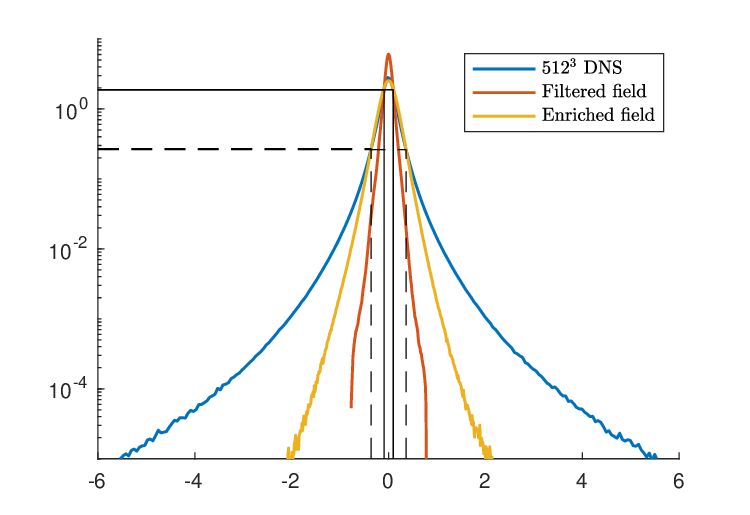}}
	\subfigure[Linear scale]{\includegraphics[trim={0.75cm 0 0.75cm 0},clip,width=.45\textwidth]{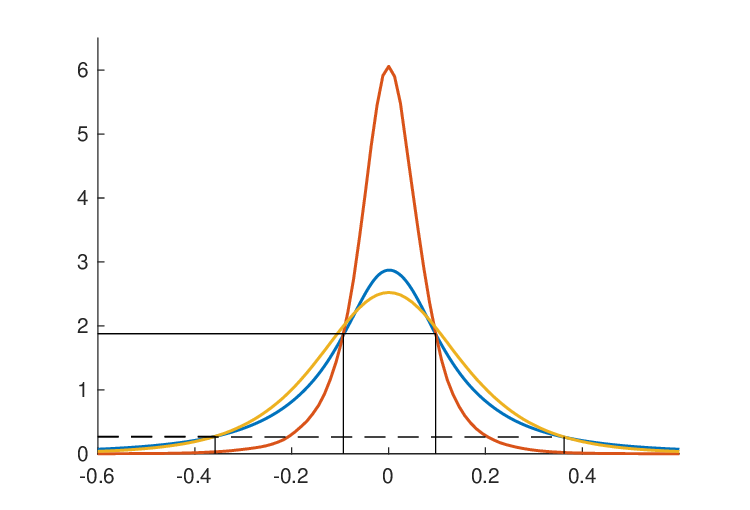}}
    \caption{Probability density functions for $y$-derivative of pressure. See previous PDF plots for description of solid/dashed black lines.}
    \label{fig:dPdy PDFs DNS512 a priori}
\end{figure}

The pressure spectra in Fig. (\ref{fig:Pressure spectra DNS512 a priori}) shows that the DNS resolution is insufficient to capture an inertial subrange but the Gabor induced-field still does a good job extrapolating to smaller scales. Improvement to the pressure and pressure derivative PDFs (Figs.  [\ref{fig:P PDFs DNS512 a priori}] and [\ref{fig:dPdy PDFs DNS512 a priori}]) is similar to the inviscid results in that Pressure events with 32.42\% probability of occurring are now captured in the enriched field which were absent in the large scales and recovery of events with 42.06\% probability of occurring for the pressure derivative.

Additionally, similar to that seen in the LES discussed above, the pressure gradient spectra show moderate agreement with the benchmark in terms of high wavenumber content, which is a significant improvement over the filtered field by itself.

\begin{table}
\centering
    \begin{tabular}{|c|c|c|c|c|}
    \hline
       Case  &  Var($P$) & Error(\%) & Var($\partial P/ \partial y$) & Error(\%) \\ \hline
       $512^3$ DNS (baseline) & 0.0015 & 0 & 0.0743 & 0\\ \hline
        F, $k_{co} = (2/3)16$ & 0.0012 & 20.11 & 0.0077 & 89.70\\ 
        E, $k_{co} = (2/3)16$ & 0.0014 & 6.89 & 0.0372 &  50.0 \\ 
        \hline
    \end{tabular}
    \caption{Pressure statistics and error associated with the filtered (``F'') and enriched (``E'') fields}
    \label{Pressure stats table 3}
\end{table}

\begin{figure}[htp]%[!ht]
    \setlength{\textfloatsep}{10pt plus 1.0pt minus 2.0pt}
    \setlength{\floatsep}{6pt plus 1.0pt minus 1.0pt}
    \setlength{\intextsep}{6pt plus 1.0pt minus 1.0pt}
    {\includegraphics[width=1\linewidth]{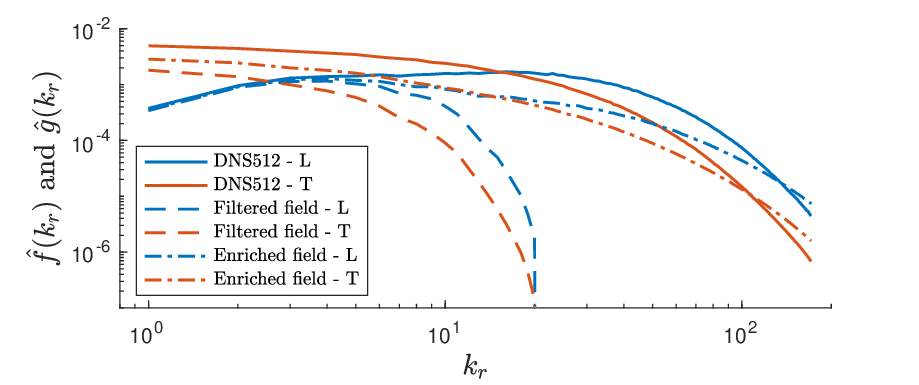}}\\
    \caption{Pressure gradient spectra}
    \label{fig:gradP spectra}
\end{figure}

\bibliography{Mybib}

\begin{thebibliography}{21}%
\makeatletter
\providecommand \@ifxundefined [1]{%
 \@ifx{#1\undefined}
}%
\providecommand \@ifnum [1]{%
 \ifnum #1\expandafter \@firstoftwo
 \else \expandafter \@secondoftwo
 \fi
}%
\providecommand \@ifx [1]{%
 \ifx #1\expandafter \@firstoftwo
 \else \expandafter \@secondoftwo
 \fi
}%
\providecommand \natexlab [1]{#1}%
\providecommand \enquote  [1]{``#1''}%
\providecommand \bibnamefont  [1]{#1}%
\providecommand \bibfnamefont [1]{#1}%
\providecommand \citenamefont [1]{#1}%
\providecommand \href@noop [0]{\@secondoftwo}%
\providecommand \href [0]{\begingroup \@sanitize@url \@href}%
\providecommand \@href[1]{\@@startlink{#1}\@@href}%
\providecommand \@@href[1]{\endgroup#1\@@endlink}%
\providecommand \@sanitize@url [0]{\catcode `\\12\catcode `\$12\catcode
  `\&12\catcode `\#12\catcode `\^12\catcode `\_12\catcode `\%12\relax}%
\providecommand \@@startlink[1]{}%
\providecommand \@@endlink[0]{}%
\providecommand \url  [0]{\begingroup\@sanitize@url \@url }%
\providecommand \@url [1]{\endgroup\@href {#1}{\urlprefix }}%
\providecommand \urlprefix  [0]{URL }%
\providecommand \Eprint [0]{\href }%
\providecommand \doibase [0]{https://doi.org/}%
\providecommand \selectlanguage [0]{\@gobble}%
\providecommand \bibinfo  [0]{\@secondoftwo}%
\providecommand \bibfield  [0]{\@secondoftwo}%
\providecommand \translation [1]{[#1]}%
\providecommand \BibitemOpen [0]{}%
\providecommand \bibitemStop [0]{}%
\providecommand \bibitemNoStop [0]{.\EOS\space}%
\providecommand \EOS [0]{\spacefactor3000\relax}%
\providecommand \BibitemShut  [1]{\csname bibitem#1\endcsname}%
\let\auto@bib@innerbib\@empty
%</preamble>
\bibitem [{\citenamefont {Ghate}(2018)}]{GhateThesis}%
  \BibitemOpen
  \bibfield  {author} {\bibinfo {author} {\bibfnamefont {A.~S.}\ \bibnamefont
  {Ghate}},\ }\emph {\bibinfo {title} {Gabor mode enrichment in large eddy
  simulation of turbulent flows.}},\ \href@noop {} {Ph.D. thesis},\ \bibinfo
  {school} {Stanford University} (\bibinfo {year} {2018})\BibitemShut {NoStop}%
\bibitem [{\citenamefont {Slotnick}\ \emph {et~al.}(2014)\citenamefont
  {Slotnick}, \citenamefont {Khodadoust}, \citenamefont {Alonso}, \citenamefont
  {Darmofal}, \citenamefont {Gropp}, \citenamefont {Lurie},\ and\ \citenamefont
  {Mavriplis}}]{slotnick2014cfd}%
  \BibitemOpen
  \bibfield  {author} {\bibinfo {author} {\bibfnamefont {J.}~\bibnamefont
  {Slotnick}}, \bibinfo {author} {\bibfnamefont {A.}~\bibnamefont
  {Khodadoust}}, \bibinfo {author} {\bibfnamefont {J.}~\bibnamefont {Alonso}},
  \bibinfo {author} {\bibfnamefont {D.}~\bibnamefont {Darmofal}}, \bibinfo
  {author} {\bibfnamefont {W.}~\bibnamefont {Gropp}}, \bibinfo {author}
  {\bibfnamefont {E.}~\bibnamefont {Lurie}},\ and\ \bibinfo {author}
  {\bibfnamefont {D.}~\bibnamefont {Mavriplis}},\ }\href@noop {} {\emph
  {\bibinfo {title} {CFD vision 2030 study: a path to revolutionary
  computational aerosciences}}},\ \bibinfo {type} {Tech. Rep.}\ \bibinfo
  {number} {NASA CR-2014-218178}\ (\bibinfo {year} {2014})\BibitemShut
  {NoStop}%
\bibitem [{\citenamefont {Ghate}\ \emph
  {et~al.}(2020{\natexlab{a}})\citenamefont {Ghate}, \citenamefont {Housman},
  \citenamefont {Stich}, \citenamefont {Kenway},\ and\ \citenamefont
  {Kiris}}]{ghate2020scale}%
  \BibitemOpen
  \bibfield  {author} {\bibinfo {author} {\bibfnamefont {A.~S.}\ \bibnamefont
  {Ghate}}, \bibinfo {author} {\bibfnamefont {J.~A.}\ \bibnamefont {Housman}},
  \bibinfo {author} {\bibfnamefont {G.-D.}\ \bibnamefont {Stich}}, \bibinfo
  {author} {\bibfnamefont {G.}~\bibnamefont {Kenway}},\ and\ \bibinfo {author}
  {\bibfnamefont {C.~C.}\ \bibnamefont {Kiris}},\ }\bibfield  {title} {\bibinfo
  {title} {Scale resolving simulations of the nasa juncture flow model using
  the lava solver},\ }in\ \href@noop {} {\emph {\bibinfo {booktitle} {AIAA
  Aviation 2020 Forum}}}\ (\bibinfo {year} {2020})\ p.\ \bibinfo {pages}
  {2735}\BibitemShut {NoStop}%
\bibitem [{\citenamefont {Spalart}(2015)}]{spalart2015philosophies}%
  \BibitemOpen
  \bibfield  {author} {\bibinfo {author} {\bibfnamefont {P.~R.}\ \bibnamefont
  {Spalart}},\ }\bibfield  {title} {\bibinfo {title} {Philosophies and
  fallacies in turbulence modeling},\ }\href@noop {} {\bibfield  {journal}
  {\bibinfo  {journal} {Progress in Aerospace Sciences}\ }\textbf {\bibinfo
  {volume} {74}},\ \bibinfo {pages} {1} (\bibinfo {year} {2015})}\BibitemShut
  {NoStop}%
\bibitem [{\citenamefont {Ghate}\ and\ \citenamefont
  {Lele}(2017)}]{GhateJFM2017}%
  \BibitemOpen
  \bibfield  {author} {\bibinfo {author} {\bibfnamefont {A.~S.}\ \bibnamefont
  {Ghate}}\ and\ \bibinfo {author} {\bibfnamefont {S.~K.}\ \bibnamefont
  {Lele}},\ }\bibfield  {title} {\bibinfo {title} {Subfilter-scale enrichment
  of planetary boundary layer large eddy simulation using discrete
  fourier-gabor modes},\ }\href@noop {} {\bibfield  {journal} {\bibinfo
  {journal} {Journal of Fluid Mechanics}\ }\textbf {\bibinfo {volume} {819}},\
  \bibinfo {pages} {494} (\bibinfo {year} {2017})}\BibitemShut {NoStop}%
\bibitem [{\citenamefont {Batchelor}(1953)}]{batchelor1953theory}%
  \BibitemOpen
  \bibfield  {author} {\bibinfo {author} {\bibfnamefont {G.~K.}\ \bibnamefont
  {Batchelor}},\ }\href@noop {} {\emph {\bibinfo {title} {The theory of
  homogeneous turbulence}}}\ (\bibinfo  {publisher} {Cambridge university
  press},\ \bibinfo {year} {1953})\BibitemShut {NoStop}%
\bibitem [{\citenamefont {Ghate}\ \emph
  {et~al.}(2020{\natexlab{b}})\citenamefont {Ghate}, \citenamefont {Towne},\
  and\ \citenamefont {Lele}}]{ghate2020broadband}%
  \BibitemOpen
  \bibfield  {author} {\bibinfo {author} {\bibfnamefont {A.}~\bibnamefont
  {Ghate}}, \bibinfo {author} {\bibfnamefont {A.}~\bibnamefont {Towne}},\ and\
  \bibinfo {author} {\bibfnamefont {S.}~\bibnamefont {Lele}},\ }\bibfield
  {title} {\bibinfo {title} {Broadband reconstruction of inhomogeneous
  turbulence using spectral proper orthogonal decomposition and gabor modes},\
  }\href@noop {} {\bibfield  {journal} {\bibinfo  {journal} {Journal of Fluid
  Mechanics}\ }\textbf {\bibinfo {volume} {888}} (\bibinfo {year}
  {2020}{\natexlab{b}})}\BibitemShut {NoStop}%
\bibitem [{\citenamefont {Fung}\ \emph {et~al.}(1992)\citenamefont {Fung},
  \citenamefont {Hunt}, \citenamefont {Malik},\ and\ \citenamefont
  {Perkins}}]{fung1992kinematic}%
  \BibitemOpen
  \bibfield  {author} {\bibinfo {author} {\bibfnamefont {J.~C.~H.}\
  \bibnamefont {Fung}}, \bibinfo {author} {\bibfnamefont {J.~C.}\ \bibnamefont
  {Hunt}}, \bibinfo {author} {\bibfnamefont {N.}~\bibnamefont {Malik}},\ and\
  \bibinfo {author} {\bibfnamefont {R.}~\bibnamefont {Perkins}},\ }\bibfield
  {title} {\bibinfo {title} {Kinematic simulation of homogeneous turbulence by
  unsteady random fourier modes},\ }\href@noop {} {\bibfield  {journal}
  {\bibinfo  {journal} {Journal of Fluid Mechanics}\ }\textbf {\bibinfo
  {volume} {236}},\ \bibinfo {pages} {281} (\bibinfo {year}
  {1992})}\BibitemShut {NoStop}%
\bibitem [{\citenamefont {Mann}(1994)}]{mann1994spatial}%
  \BibitemOpen
  \bibfield  {author} {\bibinfo {author} {\bibfnamefont {J.}~\bibnamefont
  {Mann}},\ }\bibfield  {title} {\bibinfo {title} {The spatial structure of
  neutral atmospheric surface-layer turbulence},\ }\href@noop {} {\bibfield
  {journal} {\bibinfo  {journal} {Journal of fluid mechanics}\ }\textbf
  {\bibinfo {volume} {273}},\ \bibinfo {pages} {141} (\bibinfo {year}
  {1994})}\BibitemShut {NoStop}%
\bibitem [{\citenamefont {Ghate}\ and\ \citenamefont
  {Lele}(2020)}]{ghate_lele_2020}%
  \BibitemOpen
  \bibfield  {author} {\bibinfo {author} {\bibfnamefont {A.~S.}\ \bibnamefont
  {Ghate}}\ and\ \bibinfo {author} {\bibfnamefont {S.~K.}\ \bibnamefont
  {Lele}},\ }\bibfield  {title} {\bibinfo {title} {Gabor mode enrichment in
  large eddy simulations of turbulent flow},\ }\href@noop {} {\bibfield
  {journal} {\bibinfo  {journal} {Journal of Fluid Mechanics}\ }\textbf
  {\bibinfo {volume} {903}},\ \bibinfo {pages} {A13} (\bibinfo {year}
  {2020})}\BibitemShut {NoStop}%
\bibitem [{\citenamefont {He}\ \emph {et~al.}(2017)\citenamefont {He},
  \citenamefont {Jin},\ and\ \citenamefont {Yang}}]{he2017space}%
  \BibitemOpen
  \bibfield  {author} {\bibinfo {author} {\bibfnamefont {G.}~\bibnamefont
  {He}}, \bibinfo {author} {\bibfnamefont {G.}~\bibnamefont {Jin}},\ and\
  \bibinfo {author} {\bibfnamefont {Y.}~\bibnamefont {Yang}},\ }\bibfield
  {title} {\bibinfo {title} {Space-time correlations and dynamic coupling in
  turbulent flows},\ }\href@noop {} {\bibfield  {journal} {\bibinfo  {journal}
  {Annual Review of Fluid Mechanics}\ }\textbf {\bibinfo {volume} {49}},\
  \bibinfo {pages} {51} (\bibinfo {year} {2017})}\BibitemShut {NoStop}%
\bibitem [{\citenamefont {Nicoud}\ \emph {et~al.}(2011)\citenamefont {Nicoud},
  \citenamefont {Toda}, \citenamefont {Cabrit}, \citenamefont {Bose},\ and\
  \citenamefont {Lee}}]{nicoud2011using}%
  \BibitemOpen
  \bibfield  {author} {\bibinfo {author} {\bibfnamefont {F.}~\bibnamefont
  {Nicoud}}, \bibinfo {author} {\bibfnamefont {H.~B.}\ \bibnamefont {Toda}},
  \bibinfo {author} {\bibfnamefont {O.}~\bibnamefont {Cabrit}}, \bibinfo
  {author} {\bibfnamefont {S.}~\bibnamefont {Bose}},\ and\ \bibinfo {author}
  {\bibfnamefont {J.}~\bibnamefont {Lee}},\ }\bibfield  {title} {\bibinfo
  {title} {Using singular values to build a subgrid-scale model for large eddy
  simulations},\ }\href@noop {} {\bibfield  {journal} {\bibinfo  {journal}
  {Physics of fluids}\ }\textbf {\bibinfo {volume} {23}},\ \bibinfo {pages}
  {085106} (\bibinfo {year} {2011})}\BibitemShut {NoStop}%
\bibitem [{\citenamefont {Carati}\ \emph {et~al.}(1995)\citenamefont {Carati},
  \citenamefont {Ghosal},\ and\ \citenamefont
  {Moin}}]{carati1995representation}%
  \BibitemOpen
  \bibfield  {author} {\bibinfo {author} {\bibfnamefont {D.}~\bibnamefont
  {Carati}}, \bibinfo {author} {\bibfnamefont {S.}~\bibnamefont {Ghosal}},\
  and\ \bibinfo {author} {\bibfnamefont {P.}~\bibnamefont {Moin}},\ }\bibfield
  {title} {\bibinfo {title} {On the representation of backscatter in dynamic
  localization models},\ }\href@noop {} {\bibfield  {journal} {\bibinfo
  {journal} {Physics of Fluids}\ }\textbf {\bibinfo {volume} {7}},\ \bibinfo
  {pages} {606} (\bibinfo {year} {1995})}\BibitemShut {NoStop}%
\bibitem [{\citenamefont {Gotoh}\ and\ \citenamefont
  {Fukayama}(2001)}]{GotohPhysRev2001}%
  \BibitemOpen
  \bibfield  {author} {\bibinfo {author} {\bibfnamefont {T.}~\bibnamefont
  {Gotoh}}\ and\ \bibinfo {author} {\bibfnamefont {D.}~\bibnamefont
  {Fukayama}},\ }\bibfield  {title} {\bibinfo {title} {Pressure spectrum in
  homogeneous turbulence},\ }\href@noop {} {\bibfield  {journal} {\bibinfo
  {journal} {Physical Review Letters}\ }\textbf {\bibinfo {volume} {86}},\
  \bibinfo {pages} {3775} (\bibinfo {year} {2001})}\BibitemShut {NoStop}%
\bibitem [{\citenamefont {Monin}\ and\ \citenamefont
  {Yaglom}(2013)}]{monin2013statistical}%
  \BibitemOpen
  \bibfield  {author} {\bibinfo {author} {\bibfnamefont {A.~S.}\ \bibnamefont
  {Monin}}\ and\ \bibinfo {author} {\bibfnamefont {A.~M.}\ \bibnamefont
  {Yaglom}},\ }\href@noop {} {\emph {\bibinfo {title} {Statistical fluid
  mechanics, volume II: mechanics of turbulence}}},\ Vol.~\bibinfo {volume}
  {2}\ (\bibinfo  {publisher} {Courier Corporation},\ \bibinfo {year}
  {2013})\BibitemShut {NoStop}%
\bibitem [{\citenamefont {George}\ \emph {et~al.}(1984)\citenamefont {George},
  \citenamefont {Beuther},\ and\ \citenamefont {Arndt}}]{george1984pressure}%
  \BibitemOpen
  \bibfield  {author} {\bibinfo {author} {\bibfnamefont {W.~K.}\ \bibnamefont
  {George}}, \bibinfo {author} {\bibfnamefont {P.~D.}\ \bibnamefont
  {Beuther}},\ and\ \bibinfo {author} {\bibfnamefont {R.~E.}\ \bibnamefont
  {Arndt}},\ }\bibfield  {title} {\bibinfo {title} {Pressure spectra in
  turbulent free shear flows},\ }\href@noop {} {\bibfield  {journal} {\bibinfo
  {journal} {J. Fluid Mech}\ }\textbf {\bibinfo {volume} {148}},\ \bibinfo
  {pages} {155} (\bibinfo {year} {1984})}\BibitemShut {NoStop}%
\bibitem [{Note1()}]{Note1}%
  \BibitemOpen
  \bibinfo {note} {The pressure gradient is here represented as only the
  $y$-derivative of pressure. This being isotropic turbulence the other
  components are indistinguishable.}\BibitemShut {Stop}%
\bibitem [{Note2()}]{Note2}%
  \BibitemOpen
  \bibinfo {note} {It should be noted that the extremely large improvement in
  pressure variance is a bit misleading since these are instantaneous
  quantities and change from time-step to time-step. That said, the pressure is
  well represented by the LES and it is the pressure derivative that benefits
  the most from the enrichment procedure. This is evident by visually comparing
  the PDFs in Figures (\ref {fig:Pressure PDFs a posteriori}) \& (\ref
  {fig:dPdy PDFs a posteriori})}\BibitemShut {NoStop}%
\bibitem [{\citenamefont {Park}\ and\ \citenamefont
  {Moin}(2016)}]{park2016space}%
  \BibitemOpen
  \bibfield  {author} {\bibinfo {author} {\bibfnamefont {G.~I.}\ \bibnamefont
  {Park}}\ and\ \bibinfo {author} {\bibfnamefont {P.}~\bibnamefont {Moin}},\
  }\bibfield  {title} {\bibinfo {title} {Space-time characteristics of
  wall-pressure and wall shear-stress fluctuations in wall-modeled large eddy
  simulation},\ }\href@noop {} {\bibfield  {journal} {\bibinfo  {journal}
  {Physical review fluids}\ }\textbf {\bibinfo {volume} {1}},\ \bibinfo {pages}
  {024404} (\bibinfo {year} {2016})}\BibitemShut {NoStop}%
\bibitem [{\citenamefont {Bose}\ and\ \citenamefont
  {Park}(2018)}]{bose2018wall}%
  \BibitemOpen
  \bibfield  {author} {\bibinfo {author} {\bibfnamefont {S.~T.}\ \bibnamefont
  {Bose}}\ and\ \bibinfo {author} {\bibfnamefont {G.~I.}\ \bibnamefont
  {Park}},\ }\bibfield  {title} {\bibinfo {title} {Wall-modeled large-eddy
  simulation for complex turbulent flows},\ }\href@noop {} {\bibfield
  {journal} {\bibinfo  {journal} {Annual review of fluid mechanics}\ }\textbf
  {\bibinfo {volume} {50}},\ \bibinfo {pages} {535} (\bibinfo {year}
  {2018})}\BibitemShut {NoStop}%
\bibitem [{\citenamefont {Pope}(2001)}]{pope2001turbulent}%
  \BibitemOpen
  \bibfield  {author} {\bibinfo {author} {\bibfnamefont {S.~B.}\ \bibnamefont
  {Pope}},\ }\href@noop {} {\bibinfo {title} {Turbulent flows}} (\bibinfo
  {year} {2001})\BibitemShut {NoStop}%
\end{thebibliography}%
% Produces the bibliography via BibTeX.

\end{document}